\documentclass[journal]{IEEEtran}
\usepackage{xcolor,soul,framed} 

\colorlet{shadecolor}{yellow}
\usepackage[pdftex]{graphicx}
\graphicspath{{../pdf/}{../jpeg/}}
\DeclareGraphicsExtensions{.pdf,.jpeg,.png}

\usepackage[cmex10]{amsmath}
\usepackage{bm}
\usepackage{amssymb}
\usepackage{gensymb}
\usepackage{array}
\usepackage{mdwmath}
\usepackage{mdwtab}
\usepackage{eqparbox}
\usepackage{url}
\usepackage{algorithm}  
\usepackage{algpseudocode}
\usepackage{tabularx}
\usepackage{tabularray}
\usepackage{tablefootnote}
\usepackage{threeparttable}
\usepackage{subfigure}
\usepackage{booktabs}
\usepackage{multicol}
\usepackage{multirow}
\usepackage{makecell}
\usepackage{setspace}

\newtheorem{theorem}{Theorem}

\newtheorem{corollary}{Corollary}

\begin{document}
\bstctlcite{IEEEexample:BSTcontrol}
    \title{Power System Robust State Estimation As a Layer: An Optimization-embedded End-to-end \\ Learning Approach}
  \author{Yibo~Ding,~\IEEEmembership{Student Member,~IEEE,}
      Wenzhuo~Shi,~\IEEEmembership{Student Member,~IEEE,} \\
      Mengzhao~Duan,~\IEEEmembership{Student Member,~IEEE,}
      Yuhong~Zhao,~\IEEEmembership{Student Member,~IEEE,}\\
      Jiaqi~Ruan,~\IEEEmembership{Member,~IEEE,}
      Jian~Zhao,~\IEEEmembership{Member,~IEEE,}
      and~Zhao~Xu,~\IEEEmembership{Senior Member,~IEEE}}



\maketitle

\begin{abstract}
Serving as an essential prerequisite for modern power system operation, robust state estimation (RSE) could effectively resist noises and outliers in measurements. The emerging neural network (NN) based end-to-end (E2E) learning framework enables real-time application of RSE but potentially yields solutions that are statistically accurate yet physically inconsistent. To bridge this gap, this work proposes a novel E2E learning based RSE framework, where the convex-relaxed RSE problem is innovatively constructed as an explicit differentiable layer into an NN as the first trial. This optimization-embedded layer (termed as ‘Opt-Layer’ in our work) serves as a solver of the RSE problem. Then, the relaxed solutions are recovered through post-processing layers. Through seamlessly embedding the underlying KKT conditions into the gradients during backward propagation, the physical consistency in the estimated states could be significantly enhanced, realizing lower measurement residuals. Also, the measurement weights are treated as learnable parameters of NN to enhance estimation robustness, enabling the Opt-Layer to actively denoise. A hybrid loss function is formulated to pursue accurate and physically consistent solutions. Extensive simulations have been carried out to demonstrate that the proposed framework can significantly improve the SE performance especially in terms of physical consistency on eight test systems, in comparison to classical E2E learning models, physics-informed NN (PINN) models, graph-based learning models, and conventional optimization-based approaches. The estimation performances under partial observability, severe noise contamination are systematically evaluated. Computational complexity and runtime analysis are also comprehensively demonstrated.
\end{abstract}


\IEEEpeerreviewmaketitle

\section{Introduction}
\IEEEPARstart{T}{he} ultimate goal of power system state estimation (PSSE) is to calculate state variables according to real-time measurements by minimizing the measurement errors, also called residuals, providing a snapshot of the grid's conditions for system operation and control. Traditionally, it is conducted after topology identification and observability checking, and followed by the processing of bad measurement data, namely the outliers \cite{abur2004power}. Real-time measurements are often available from supervisory control and data acquisition (SCADA) systems in modern smart grids \cite{cheng2023survey}. The measurement set typically comprises bus voltage magnitudes, nodal power injections, and branch power flows. The state variables to be estimated are the complex voltages, i.e., the voltage magnitudes and phase angles at all buses.

Conventionally, PSSE is formulated as a constrained optimization problem calculating the weighted least squares (WLS) of residuals under the Gaussian assumption of noises, solved by iterative approaches such as Newton-Raphson algorithm \cite{minot2015distributed}. Nevertheless, the WLS-based estimator suffers from sensitivity in the presence of outliers originating from sensor malfunctions or communication failures \cite{liu2022robust}. Outliers would therefore introduce a significant bias to the estimation results. Traditionally, the chi-square test is carried out to identify outliers, followed by the corresponding removal and repeated estimation \cite{grainger1999power}. To enhance the estimation robustness against outliers, research has shifted towards robust state estimation (RSE). One prominent RSE approach is to dynamically adjust the measurement weights \cite{zhao2015power} \cite{wang2019robust}, which are assumed as constants in classic PSSE models \cite{abur2004power}. To be specific, the weights are iteratively updated according to the residuals calculated in the repeated estimation \cite{pires1999iteratively} \cite{zhong2004auto}, or optimized through specially training a neural network (NN) during measurements dataset preprocessing \cite{patel2023weighted}. Another approach is to adopt the weighted least absolute values (WLAV)-based estimator, which inherently mitigates the influence of outliers, eliminating the need for the processing of outliers and the time-consuming repeated estimation \cite{celik2002robust}\cite{gol2014lav}.

In recent years, data-driven approaches, particularly end-to-end (E2E) learning, have the potential to serve as a holistic framework with superior performance to handle topology identification, PSSE and outliers processing altogether, provided adequate representative data are available for training \cite{wu2022gridtopo}. For the RSE problem in this work, the E2E learning is naturally credited by its robustness to non-Gaussian noise and outliers, and the near-instantaneous output in the online application, i.e. an NN can be trained offline to directly map noisy measurements to state variables \cite{9758816}. However, the classical E2E learning approaches struggle to handle the power flow based physical constraints involved in the RSE problem, since it was initially developed for non-constrained problems \cite{donti2017task}. Consequently, the resulted estimation may achieve high regression accuracy but is not guaranteed to be physically consistent \cite{ding2025optimal}. In the context of RSE, enhanced physical consistency indicates that the estimated states are closer to the true physical states of the power system, quantified by measurement residuals in the objective.

To bridge this gap, one possible solution is to embed physical constraints into the E2E learning framework. A typical implementation involves designing a physics-informed NN (PINN) by encoding physical constraints into a complicated and hybrid loss function \cite{falas2025robust} \cite{ngo2024physics} \cite{9072507}. As such, the PINN is just expected to generate solutions that are not only accurate but also physically consistent. In principle, the PINN approach may not generate the genuine optimal solution, since the Karush-Kuhn-Tucker (KKT) optimality conditions are not explicitly fulfilled \cite{boyd2004convex}. Specifically, standard PINNs typically treat physical constraints as external, soft penalties added to the loss function. This penalty-based approach merely encourages the outputs of NN to move towards the feasible region, but it lacks the structural mechanism to guarantee physical consistency.

Alternatively, embedding an independent differentiable optimization layer constructed from the optimization problem itself into the NN structure has become a promising solution \cite{amos2017optnet}. Once trained properly, the outputs of such NN can largely improve the alignment with physical constraints during online application, since the derivatives of the loss function with respect to problem parameters for backward propagation (BP) during training is derived from the KKT conditions \cite{zhu2025decision}. Compared to the PINN, where the physical constraints are exogenously added into the loss function only, such an optimization-embedded framework can thus significantly strengthen the constraint alignments. This novel framework has been applied to AC optimal power flow (OPF) \cite{jia2024optnet}\cite{xie2025neural} and voltage regulation \cite{sang2022safety}, but has never been used in PSSE by far.

In this work, a novel E2E learning framework embedded with the differentiable optimization layer dedicated for RSE is proposed. Trained under a hybrid loss function gauging SE performance and physical consistency, the proposed framework is capable of estimating system states with significantly lowered residuals. Also, to enable the NN to actively identify the outliers for more robustness, the measurement weights are treated as learnable parameters, which still remains notably underrepresented in the existing literature. 

A primary obstacle to implementing the aforementioned framework is the non-convexity of AC power flow equations as the constraints of the RSE problem \cite{agrawal2019differentiable}. Therefore, convex relaxation becomes necessary to derive the KKT conditions of the global optimal solution and then calculate the derivatives. Some relaxation techniques introduce auxiliary variables to linearize the power flow equations and impose second-order conic constraints on these variables, reformulating the problem as a second-order cone programming (SOCP) \cite{chen2020robust} \cite{aghamolki2018socp}. However, a significant drawback is that the exactness can only be guaranteed for power grids with radial topology. The semidefinite relaxation is thus proposed for grids with either radial or meshed topologies \cite{zhu2014power}, where the introduced auxiliary variables should satisfy a semidefinite constraint and a non-convex rank-1 constraint. This relaxation is usually exact, especially when the rank-1 solution can be obtained \cite{vandenberghe1996semidefinite}. While methods such as rank reduction and convex iteration have been employed to tackle with the non-convexity of rank-1 constraint \cite{yao2018distribution}, computational burdens arising from the semidefinite constraint exist. Correspondingly, gradient-based acceleration strategies \cite{lan2019fast} and conic relaxation \cite{zhang2017conic} have been proposed to boost the tractability of semidefinite-relaxed problem. The semidefinite relaxation with a regularization term added into the RSE objective is utilized in our proposed E2E learning framework. The regularization term is designed to implicitly guide towards the rank-1 solution. 

The major contributions are summarized as follows. Firstly, a novel optimization-embedded E2E learning framework is proposed for RSE. To the authors’ best knowledge, this work is the first to both formulate the RSE problem as an independent differentiable layer within a NN and treat measurement weights as learnable parameters, enabling the NN to actively filter outliers towards higher estimation robustness in a holistic manner. Unlike PINN that exogenously incorporating physical inconsistency as soft penalties, the proposed framework significantly improves overall estimation performances, as demonstrated through extensive simulations. Although the customized hybrid loss function weighing estimation accuracy and physical consistency is similar to that of PINN, a key advantage of the proposed framework lies in the rigorous KKT conditions embedding during BP, achieving significantly enhanced physical consistency in estimated states. To realize this, differentiation of the relaxed RSE problem is specially carried out, where homogenization and implicit function theorem are utilized.


The proposed E2E learning framework is introduced in Section \ref{sec2}. The RSE problem and the convex relaxation process are modeled in Section \ref{sec3}. The loss function design, RSE problem differentiation, and model training process are presented in Section \ref{sec4}. Section \ref{sec5} analyzes the extensive simulation results, and Section \ref{sec6} draws the conclusion.

\section{Framework Overview}\label{sec2}
\begin{figure}
    \centering
    \includegraphics[width=0.95\linewidth]{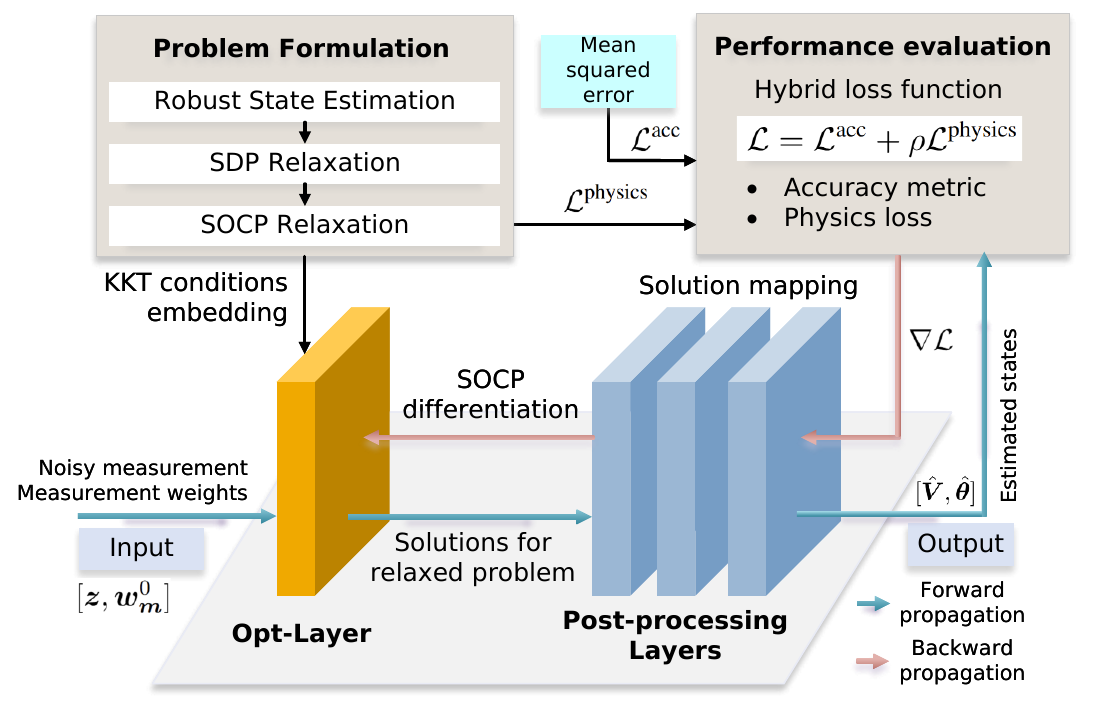}
    \caption{The proposed optimization-embedded end-to-end learning framework for power system RSE.}
    \label{fig:Framework}
\end{figure}

Figure \ref{fig:Framework} demonstrates the proposed E2E learning framework for power system RSE problem. The input context of the NN includes noisy measurement data and the initial value of learnable weights of measurements. After offline training, the E2E learning framework can directly output the estimated states given real-time measurements in online implementation.

The NN in the proposed E2E framework is constructed by a differentiable optimization layer, named as '$\mathrm{Opt-Layer}$', and a set of post-processing layers. The $\mathrm{Opt-Layer}$ is rigorously and inherently embedded with physical constraints (i.e., the power flow based measurement equations) in the RSE problem, and the post-processing layers recover the solution, namely mapping the solutions of relaxed RSE problem into the space of decision variables of the original problem. The post-processing layers could be, e.g. standard dense layers. Unlike existing differentiable optimization literature \cite{jia2024optnet}\cite{xie2025neural}\cite{sang2022safety}, our $\mathrm{Opt-Layer}$ is mathematically tailored for the RSE problem. It incorporates a physically meaningful regularization term to guide relaxation exactness, and explicitly leverages grid sparsity to ensure computational tractability, which will be introduced in Section III-B and IV-D.

To realize seamless physical constraints embedding in BP, convex relaxation of the original RSE problem is necessary to obtain the KKT conditions of the global optimum. The mathematical details will be introduced in Section \ref{sec3}. Specifically, the RSE problem is firstly relaxed into a semidefinite programming (SDP) by introducing auxiliary decision variables and the corresponding semidefinite and rank-1 constraints. Then, a regularization term is integrated into the objective to replace the non-convex rank-1 constraint, which implicitly guides towards the rank-1 solution. After that, the semidefinite constraint is further relaxed into a second-order conic constraint to reduce computational complexity. The SDP-based RSE problem is thus reformulated as a SOCP, which can be readily solved by commercial solvers. Then, through differentiating the SOCP-based RSE problem, rigorous KKT conditions embedding during BP is achieved.

During the forward propagation (FP), the NN takes the noisy measurements data and learnable measurement weights as input context. Then, the $\mathrm{Opt-Layer}$ would solve the batched relaxed RSE problems for all the samples in the training dataset, then output the estimated state variables after post-processing mapping. Note that, problem sparsification is also necessary in practical implementation for acceleration. To effectively guarantee the regression accuracy and physical consistency of the NN, a hybrid loss function $\mathcal{L}$ weighing accuracy metric and physical loss is specially introduced, where the latter is derived from the objective of RSE problem, as modeled in Section \ref{sec4}.

During the BP, the derivatives of the hybrid loss function $\nabla\mathcal{L}$ with respect to the learnable parameters of the NN are calculated. Specifically, the derivatives of optimal solutions of the relaxed problem with respect to the problem parameters are calculated by homogenizing the KKT conditions and utilizing the implicit function theorem, which will be illustrated in Section \ref{sec4}-B. The NN is then trained offline via an algorithm such as stochastic gradient descent. Upon completion of the training, the NN learns a direct mapping from noisy measurements to the estimated nodal voltage magnitudes and phase angles, facilitating real-time applications.

\section{Problem Formulation}\label{sec3}
Given a fixed topology, the grid parameters could be defined as follows. Let $g_s+jb_s$ denote the series branch admittance between buses $i$ and $j$; $b_c$ represents the line charging susceptance; $k_t$ is the turns ratio of the transformer; $g_{ij} = g_s/k_t$ and $b_{ij} = b_s/k_t$ are respectively the mutual conductance and susceptance between bus $i$ and $j$; $g_{si} = (1-k_t)g_s/k^2$ and $b_{si} = (1-k_t)b_s/k^2+b_c/2$ are respectively the shunt conductance and susceptance from bus $i$ to ground. Then, for simplicity, let $G_{ij} =-g_{ij}$, $B_{ij} =-b_{ij}$, $\hat{g}_{ij}=g_{si} + g_{ij}$ and $\hat{b}_{ij} = b_{si} + b_{ij}$. Also, let $\mathcal{N}$ and $\mathcal{N}_i$ respectively denote the sets of all buses and the adjacent buses of bus $i$. Let $N=|\mathcal{N}|$ denote the number of buses.

\subsection{Classical Robust State Estimation Problem}
Classical WLS-based PSSE provides unbiased estimates when the residuals follow a zero-mean Gaussian distribution \cite{abur2004power}. In practical implementations, there might exist outliers in measurements, making the overall distribution of residuals deviate from a zero-mean Gaussian distribution. Also, the WLS-based estimator is highly sensitive to outliers due to the quadratic function of residuals. In contrast, the WLAV-based estimator, intrinsically a L1-norm minimization of residuals, exhibits higher robustness against outliers. Accordingly, given a measurement dataset $\mathcal{M}$, a WLAV-based RSE model is formulated as follows \cite{chen2020robust}:
\begin{equation}
    \mathrm{(P1)} \quad \min_{V_i,\theta_i} \quad \sum_{m\in\mathcal{M}} |\epsilon_m|/\sigma_m
\end{equation}
subject to
\begin{equation}
    z_{V_i} = V_i + \epsilon_{V_i} 
\end{equation}
\begin{equation}
    z_{P_{ij}}=V_i^2 \hat{g}_{ij} - V_iV_j g_{ij} \cos \theta_{ij} - V_iV_j b_{ij}\sin \theta_{ij}+\epsilon_{P_{ij}} \label{eq RSE Pij}
\end{equation}
\begin{equation}
    z_{Q_{ij}}=-V_i^2 \hat{b}_{ij}  + V_iV_j b_{ij} \cos \theta_{ij} - V_iV_j g_{ij}\sin \theta_{ij}+\epsilon_{Q_{ij}}
\end{equation}
\begin{equation}
    \textstyle z_{P_i} = V_i  \sum_{j\in \mathcal{N}_i} V_j (G_{ij} \cos\theta_{ij} + B_{ij}\sin \theta_{ij}) + \epsilon_{P_i}
\end{equation}
\begin{equation}
    \textstyle z_{Q_i} = V_i  \sum_{j\in \mathcal{N}_i} V_j (G_{ij} \sin\theta_{ij} - B_{ij}\cos \theta_{ij}) + \epsilon_{Q_i}  \label{eq RSE Qi}
\end{equation}
where $z_{\cdot}$ and $\epsilon_{\cdot}$ are respectively the measurement data and residuals. $\sigma_m$ is the weight of the $m$-th measurements data, where $m\in\{V_i,P_{ij},Q_{ij},P_i,Q_i\}$. $V_i$ is the voltage magnitude of bus $i$. $\theta_{ij}$ is the phase angle difference between bus $i$ and $j$. $P_{ij}$ and $Q_{ij}$ are respectively the active and reactive line power flows between bus $i$ and $j$. $P_i$ and $Q_i$ are respectively the total active and reactive power injections of bus $i$.

\textit{Remark:} The WLAV-based estimator yields unbiased estimates when the residuals follow a Laplace distribution \cite{singh2009choice}, which has fatter tails than standard Gaussian distribution, providing inherent robustness against almost inevitable outliers. Although the practical statistical distribution of residuals may not strictly follow a Laplace distribution, the WLAV-based estimator still demonstrates more robust performance than the WLS-based estimator, rendering it suitable for applications.

\subsection{Convex Relaxation}
Although $\mathrm{(P1)}$ effectively suppresses the impact of outliers, it introduces undesired non-convexity because of the quadratic voltage magnitudes and the products of voltage magnitudes with trigonometric functions of phase angles in the constraints (\ref{eq RSE Pij})-(\ref{eq RSE Qi}). To alleviate the computational burden and pave the way for rigorous physical constraint embedding, convex relaxation for $\mathrm{(P1)}$ is necessary. 

The relaxation is achieved by introducing an auxiliary complex matrix $\bm{X}=\bm{V}\bm{V}^{*} \in \mathbb{C}^{|\mathcal{N}|\times|\mathcal{N}|}$ to linearize (\ref{eq RSE Pij})-(\ref{eq RSE Qi}), where $\bm{V} \in \mathbb{C}^{|\mathcal{N}|}$ is the complex voltage vector and $\bm{V}_i=V_i\angle\theta_i$. $\bm{V}^{*}$ is the conjugate of $\bm{V}$. The diagonal elements of $\bm{X}$ are given by $X_{ii}=V_i^2$, while the off-diagonal elements are $X_{ij}=V_iV_j(\cos \theta_{ij}+j\sin \theta_{ij})$.

\begin{theorem}
    \textit{The equality $\bm{X}=\bm{V}\bm{V}^{*}$ holds if and only if the following two conditions are met: (1) $\bm{X}\succeq \bm{0}$; and (2) $\mathrm{rank}(\bm{X})=1$} \cite{vandenberghe1996semidefinite}.
\end{theorem}

Condition (1) can be readily enforced by adding a positive semidefinite constraint. Nevertheless, Condition (2), $\mathrm{rank}(\bm{X})=1$, is still non-convex and poses computational challenge. Therefore, a regularization term is introduced into the objective to guide towards the rank-1 solution. 

\begin{corollary}
    \textit{The introduced regularization term exactly calculates the total active power loss of the power system. A proof is provided in Appendix B.}
\end{corollary}

The resulting problem $\mathrm{(P2)}$ is formulated as follows:
\begin{equation}
    \mathrm{(P2)} \ \min_{\bm{X}} \quad \rho_r\sum_{m\in\mathcal{M}} |\epsilon_m|/\sigma_m + \mathrm{Re}(\sum_{i\in \mathcal{N}} \sum_{j\in \mathcal{N}} Y_{ij}^* X_{ij})
\end{equation}
subject to
\begin{equation}
    z_{V_i^2} = X_{ii} + \epsilon_{V_i^2} \label{eq SDP V}
\end{equation}
\begin{equation}
    z_{P_{ij}} = X_{ii} \hat{g}_{ij} - g_{ij}\mathrm{Re}(X_{ij}) - b_{ij}\mathrm{Im}(X_{ij}) + \epsilon_{P_{ij}} \label{eq SDP Pij}
\end{equation}
\begin{equation}
    z_{Q_{ij}} = -X_{ii} \hat{b}_{ij} + b_{ij}\mathrm{Re}(X_{ij}) - g_{ij}\mathrm{Im}(X_{ij}) + \epsilon_{Q_{ij}}
\end{equation}
\begin{equation}
    \textstyle z_{P_i} = \sum_{j\in \mathcal{N}_i}(G_{ij} \mathrm{Re}(X_{ij}) + B_{ij}\mathrm{Im}(X_{ij})) + \epsilon_{P_i}
\end{equation}
\begin{equation}
    \textstyle z_{Q_i} = \sum_{j\in \mathcal{N}_i} (G_{ij} \mathrm{Im}(X_{ij}) - B_{ij}\mathrm{Re}(X_{ij})) + \epsilon_{Q_i} \label{eq SDP Qi}
\end{equation}
\begin{equation}
    \bm{X} \succeq \bm{0} \label{eq SDP}
\end{equation}
where $\rho_r>0$ is the weighting coefficient. $Y_{ij}$ is the off-diagonal element in the nodal admittance matrix. $\mathrm{Re}(\cdot)$ and $\mathrm{Im}(\cdot)$ are respectively the real and imaginary part of a complex variable. The second term in the objective function is the regularization term, which is actually a linear function with respect to decision variables, namely $G_{ij}\mathrm{Re}(X_{ij}) + B_{ij}\mathrm{Im}(X_{ij})$. Constraint (\ref{eq SDP}) stands for the Condition (1).

The decision variables of $\mathrm{(P2)}$ now are composed of $X_{ii}$, $\mathrm{Re}(X_{ij})$ and $\mathrm{Im}(X_{ij})$. Also, (\ref{eq RSE Pij})-(\ref{eq RSE Qi}) are all linearized in (\ref{eq SDP Pij})-(\ref{eq SDP Qi}). $\mathrm{(P2)}$ is therefore convexified into an SDP problem. 

\subsection{Second-order Conic Relaxation}
To further alleviate the computational expense of $\mathrm{(P2)}$, the positive semidefinite constraint (\ref{eq SDP}) can be relaxed into a set of more tractable second-order cone constraints. 

\begin{theorem}
    \textit{If a matrix is positive semidefinite, all of its principal submatrices must also be positive semidefinite.}
\end{theorem}

Therefore, we do not explicitly construct the complete $n \times n$ matrix. Instead, the positive semidefinite constraint is imposed on the $2 \times 2$ submatrix for each branch $(i,j)$:
\begin{equation}
    \begin{pmatrix} X_{ii} & X_{ij} \\ X_{ij}^* & X_{jj} \end{pmatrix} \geq 0 \iff X_{ii}X_{jj} \geq \operatorname{Re} (X_{ij})^2 + \operatorname{Im} (X_{ij})^2  \label{eq SOC submatrix}
\end{equation}

The exactness (rank-1 property) is achieved when the inequality holds as an equality.

Given the above theorem, $\mathrm{(P2)}$ can be exactly reformulated as the following SOCP problem $\mathrm{(P3)}$:
\begin{equation}
    \mathrm{(P3)} \ \min_{\bm{X}}  \quad \rho_r\sum_{m\in\mathcal{M}} |\epsilon_m|/\sigma_m + \mathrm{Re}(\sum_{i\in \mathcal{N}} \sum_{j\in \mathcal{N}} Y_{ij}^* X_{ij})
\end{equation}
subject to (\ref{eq SDP V})-(\ref{eq SDP Qi}),
\begin{equation}
    ||2\mathrm{Re}(X_{ij}),\
        2\mathrm{Im}(X_{ij}), \
        X_{ii}-X_{jj}||
    _2 \leq X_{ii}+X_{jj}
\end{equation}

In the context of RSE, it is hard to exactly obtain the rank-1 solution when the measurement dataset is noisy \cite{zhang2017conic}. Nevertheless, approximated rank-1 solution would be available when satisfying certain conditions, as presented in Appendix A.

\subsection{Solution Recovery}
After solving the relaxed RSE problem $\mathrm{(P3)}$, it is necessary to recover the original estimated states $\bm{V}$ from the relaxed solutions $\bm{X}$. To quantitatively examine the relaxation exactness, the average eigenvalue ratio $\bar{\lambda}$ of all the branches is specially investigated. For the submatrix of each branch $(i,j)$ depicted in \eqref{eq SOC submatrix}, the two eigenvalues $\lambda^{1,2}_{ij}$ and the eigenvalue ratio $\hat{\lambda}_{ij}$ are calculated as follows:

\begin{equation}
    \scalebox{0.83}{$\displaystyle \lambda^{1,2}_{ij} = \frac{X_{ii} + X_{jj} \pm \sqrt{(X_{ii} - X_{jj})^2 + 4[\operatorname{Re} (X_{ij})^2 + \operatorname{Im} (X_{ij})^2]}}{2}$}
\end{equation}
\begin{equation}
\hat{\lambda}_{ij}=\lambda^1_{ij}/(\lambda^1_{ij}+\lambda^2_{ij})
\end{equation}
where $\lambda^1_{ij}$ is the larger eigenvalue. If the relaxation is exact, the smaller eigenvalue $\lambda^2_{ij}\rightarrow 0$, and this ratio $\hat{\lambda}_{ij}$ will be infinitely close to 1.0.

Then, $\bar{\lambda}$ is calculated as the average value of eigenvalue ratios of all the branches. When $\bar{\lambda}$ is sufficiently close to 1 (e.g., $\bar{\lambda}\geq 0.999$), the solutions of the relaxed problem $\mathrm{(P3)}$ would be sufficiently close to the rank-1 solution, denoted as $\bm{X}^{\diamond}$.

According to Theorem 1, $\bm{X}^{\diamond}$ can be decomposed into the outer product of $\bm{V}$ and its conjugate $\bm{V}^*$. Then, by multiplying $\bm{V}$, we have:
\begin{equation}
    \bm{X}^{\diamond}\bm{V} = \bm{V}(\bm{V}^*\bm{V})=\bm{V}||\bm{V}||^2
\end{equation}

This indicates that $\bm{V}$ is an eigenvector of $\bm{X}^{\diamond}$, with the unique non-zero eigenvalue being $||\bm{V}||^2$. Since $\bm{X}^{\diamond}$ is a Hermitian positive semidefinite matrix, an eigenvalue decomposition can be performed as:
\begin{equation}
    \bm{X}^{\diamond}= \bm{U}_{e}  \bm\Lambda  \bm{U}_{e}^*
\end{equation}
where the eigenvalue matrix $\bm\Lambda$ contains only one significant non-zero eigenvalue $\lambda=||\bm{V}||^2$ , with the corresponding eigenvector $\bm{u}=\bm{V}/||\bm{V}||$. All the other eigenvalues will be theoretically zero.

Then, the complex voltage vector $\bm{V}$ could be recovered by:
\begin{equation}
    \bm{V} = ||\bm{V}||\bm{u}= \sqrt{\lambda} \bm{u}
\end{equation}

Finally, the estimated voltage magnitude $\hat{V}$ and phase angle $\hat{\theta}$ could be exactly extracted from $\bm{V}$.

\textit{Remark:} While the SOCP relaxation exactness is mathematically guaranteed for radial power grids in \textit{noiseless} power flow calculation or OPF problems \cite{low2014convex}, the presence of measurement noise in RSE intrinsically hinders the strict exactness \cite{zhang2017conic}. Nevertheless, according to the simulation results presented in Section V, we could still obtain estimates sufficiently close to the rank-1 solution ($\bar{\lambda}\geq 0.999$) for radial power grid under the RSE context. For small-scale mesh grids, we also empirically have $\bar{\lambda}\geq 0.999$ given ideal measurement redundancy. When the redundancy and observability is not ideal and when the scale of the mesh grid grows, although it would be hard for us to empirically achieve relaxation exactness, we still have approximated rank-1 solutions with $\bar{\lambda}> 0.990$. The physical consistency of estimated states from our framework remains better than conventional PINNs. For a detailed analysis, please refer to Section \ref{sec V-observe}.

\section{Learning Process for Power System RSE}\label{sec4}
\subsection{Loss Function Design}
To obtain accurate and physically consistent estimated states through the proposed E2E learning framework, a hybrid loss function $\mathcal{L}$ is defined as follows:
\begin{equation}
    \mathcal{L} = \mathcal{L}^{\text{acc}} + \rho \mathcal{L}^{\text{physics}}
    \label{eq hybrid loss}
\end{equation}
\begin{equation}
    \mathcal{L}^{\text{physics}} = \sum_{m\in\mathcal{M}} w_m|\epsilon_m| + \mathrm{Re}(\sum_{i\in \mathcal{N}} \sum_{j\in \mathcal{N}} Y_{ij}^* X_{ij})
    \label{eq physical loss}
\end{equation}
where $\mathcal{L}^{\text{acc}}$ is a differentiable accuracy metric measuring the distance between actual true states and estimated states, such as mean squared error (MSE). $\mathcal{L}^{\text{physics}}$ represents the physical inconsistency loss, composed of the WLAV of residuals and the regularization term. $w_m$ is the learnable weight of the $m$-th measurement data. $\rho$ is a hyperparameter indicating the importance of physical loss.

Note that, the non-smoothness of $|\epsilon_m|$ in $\mathcal{L}^{\text{physics}}$ is an obstacle of differentiation. Therefore, Huber loss is introduced as follows to guarantee the smoothness \cite{zhao2016robust}. Specifically, in the neighborhood of origin $\epsilon_m\in[-\delta,\delta]$, the Huber loss is defined as the quadratic function of residuals, otherwise the loss is linear. $\delta$ should be sufficiently small to achieve an approximation to the WLAV formulation.

Moreover, it is crucial to prevent the NN from cheating by assigning negative values to the learnable weights $w_m$ to minimize the loss. To enforce the positivity of $w_m$, a mapping is applied in (\ref{eq pos}). The Softplus function is specifically selected for its advantageous properties of being smooth and continuously differentiable. Besides, to ensure strict positivity and improve numerical stability, a small positive constant $\epsilon_p$ is added.
\begin{equation}
    \mathcal{L}^{\text{huber}} = \left\{
    \begin{aligned}
        & \textstyle \sum_{m\in\mathcal{M}} \hat{w}_m(\epsilon_m)^2/2, \quad \quad \ \text{for} \ |\epsilon_m|\leq \delta, \\
        & \textstyle \sum_{m\in\mathcal{M}} \hat{w}_m \delta(|\epsilon_m|-\delta/2), \ \text{otherwise.}
    \end{aligned} \right. \label{eq huber}
\end{equation}
\begin{equation}
    \hat{w}_m = \ln(1+e^{w_m}) + \epsilon_p, \ \ \forall m\in\mathcal{M}  \label{eq pos}
\end{equation}

When $\delta \rightarrow 0$, Huber loss function degenerates into the original WLAV-based loss, ensuring it is less sensitive to outliers in noisy measurement data than the quadratic WLS-based loss. To conclude, the surrogate physical loss and hybrid loss could be defined as:
\begin{equation}
    \hat{\mathcal{L}}^{\text{physics}} = \mathcal{L}^{\text{huber}}+\mathcal{L}^{\text{reg}}
\end{equation}
\begin{equation}
    \hat{\mathcal{L}} = \mathcal{L}^{\text{acc}} + \rho\hat{\mathcal{L}}^{\text{physics}}
\end{equation}
where $\mathcal{L}^{\text{reg}}$ denotes the value of regularization term, which also calculates the total active power loss for simplicity.

\textit{Remark:} Compared to linearizing the L1-norm term in (\ref{eq physical loss}), using Huber loss function as surrogate in training would not bring about extra decision variables and constraints, avoiding extra computational burden during training. Also, Huber loss function achieves near-exact approximation when $\delta$ is sufficiently close to 0.

\subsection{Differentiating the RSE Problem}
This subsection derives the derivative of the optimal solution $\bm{X}^{\diamond}$ of the relaxed RSE problem $\mathrm{(P3)}$ with respect to the problem parameters, including grid parameters and input measurements $\bm{z}$, for BP calculation.

\subsubsection{Problem Reformulation}
Firstly, the primal and dual forms of SOCP-based RSE problem $\mathrm{(P3)}$ could be rewritten in compact form as follows for simplicity:
\begin{equation}
\begin{aligned}
& \mathrm{(Primal)} \quad  \min \ \bm{c}^\top \bm{x} & \mathrm{(Dual)} \quad \min \ \bm{b}^\top \bm{y} \\
& \quad \quad \text{s.t.} \quad \bm{A}\bm{x} + \bm{s} = \bm{b} &  \quad \text{s.t.} \quad \bm{A}^\top \bm{y} + \bm{c} = \bm0 \\
& \qquad \quad \quad \quad \bm{s} \in \mathcal{K} & \qquad \  \bm{y} \in \mathcal{K}^*
\end{aligned} \label{eq primal-dual}
\end{equation}
where $\bm{x}\in \mathbb{R}^p$ collects all the primal decision variables, namely all the $X_{ii}, \mathrm{Re}(X_{ij}), \mathrm{Im}(X_{ij})$. $\bm{y}\in \mathbb{R}^d$ integrates the dual variables. $\bm{s}$ is the primal slack variable. $\mathcal{K}\subseteq \mathbb{R}^d$ is a nonempty, closed and convex cone, whose dual cone is denoted by $\mathcal{K^*}\subseteq \mathbb{R}^d$. The optimal solution of the above problem is bound to satisfy the following KKT conditions:
\begin{equation}
    \left\{ \begin{aligned}
        & \bm{A}\bm{x} + \bm{s} = \bm{b}, \\
        & \bm{A}^\top \bm{y} + \bm{c} = \bm0, \\
        &  \bm{s} \in \mathcal{K}, \quad \bm{y} \in \mathcal{K}^*, \\
        & \bm{s}^\top \bm{y}=\bm0.
    \end{aligned} \right. \label{eq KKT}
\end{equation}

\subsubsection{Differentiation Decomposition}
In this subsection, our objective is to derive the analytical derivative of the mapping from problem parameters $(\bm{A},\bm{b},\bm{c})$ to the optimal solution $(\bm{x}^*,\bm{y}^*,\bm{s}^*)$ for BP, denoted as $\Pi:\mathbb{R}^{d\times p}\times \mathbb{R}^{d} \times \mathbb{R}^{p} \rightarrow \mathbb{R}^{p+2d}$.

The mapping $\Pi$ could be decomposed into three sub-mappings $\phi  \circ g  \circ \Omega$, where $\circ$ is the function composition operator. Specifically, let $D_1=p+d+1$ and $D_2 = p+2d+2$, $\Omega:\mathbb{R}^{d\times p}\times \mathbb{R}^{d} \times \mathbb{R}^{p} \rightarrow \mathbb{R}^{D_1\times D_2}$ maps the problem parameters to an auxiliary matrix $\Omega$. $g:\mathbb{R}^{D_1\times D_2} \rightarrow \mathbb{R}^{D_2}$ is the mapping from $\Omega$ to the solution of the homogeneous self-dual system (HSDS), which will be introduced in the next subsection. $\phi:\mathbb{R}^{D_2}\rightarrow \mathbb{R}^{p+2d}$ recovers the optimal solution of (\ref{eq primal-dual}). Then, the derivative of FP $\nabla_{\Pi}$ can be calculated as follows: 
\begin{equation}
    \nabla _{\Pi}(\bm{A},\bm{b},\bm{c})= \nabla_{\phi}(g(\Omega))\nabla_g(\Omega(\bm{A},\bm{b},\bm{c}))\nabla_\Omega(\bm{A},\bm{b},\bm{c})  \label{eq gradient forward}
\end{equation}

For BP, the resulting adjoint derivative $\nabla^\top _{\Pi}$ could be calculated as follows:
\begin{equation}
    (\frac{\partial \mathcal{L}}{\partial\bm{A}},\frac{\partial \mathcal{L}}{\partial\bm{b}},\frac{\partial \mathcal{L}}{\partial\bm{c}})=\nabla^\top _{\Pi}(\bm{A},\bm{b},\bm{c})(\frac{\partial \mathcal{L}}{\partial\bm{x}},\frac{\partial \mathcal{L}}{\partial\bm{y}},\frac{\partial \mathcal{L}}{\partial\bm{s}})
\end{equation}
\begin{equation}
    \nabla^\top _{\Pi}(\bm{A},\bm{b},\bm{c})= \nabla^\top_\Omega(\bm{A},\bm{b},\bm{c})\nabla^\top_g(\Omega(\bm{A},\bm{b},\bm{c})) \nabla^\top_{\phi}(g(\Omega))   \label{eq gradient backward}
\end{equation}

\subsubsection{Pre-processing Problem Parameters}
To calculate the analytical expressions for each component in (\ref{eq gradient forward}), the auxiliary matrix $\Omega$ is first constructed to pre-process the problem parameters $(\bm{A},\bm{b},\bm{c})$ as follows:
\begin{equation}
    \Omega=\begin{bmatrix}
            \mathbf{0} & \bm A^\top & \bm 0 & \bm c & \bm 0\\
            -\bm A & \mathbf{0} & -\bm I & \bm b & \bm 0 \\
            -\bm c^\top & -\bm b^\top & \mathbf{0} & \bm 0 & -\bm I
        \end{bmatrix}.
\end{equation}

Correspondingly, the derivative of the mapping from $(\bm{A},\bm{b},\bm{c})$ to $\Omega$ can be calculated as follows:
\begin{equation}
    \nabla_\Omega(\bm{A},\bm{b},\bm{c})=\begin{bmatrix}
            \mathbf{0} & \frac{\partial \mathcal{L}}{\partial\bm A^\top}  & \bm 0 & \frac{\partial \mathcal{L}}{\partial\bm c} & \bm 0\\
            -\frac{\partial \mathcal{L}}{\partial\bm A} & \mathbf{0} & \bm 0 & \frac{\partial \mathcal{L}}{\partial\bm b} & \bm 0 \\
            -\frac{\partial \mathcal{L}}{\partial\bm c^\top} & -\frac{\partial \mathcal{L}}{\partial\bm b^\top} & \mathbf{0} & \bm 0 & \bm 0
        \end{bmatrix}.
\end{equation}

\subsubsection{Homogeneous Self-Dual Embedding}
Then, the mapping $g$ embeds the original problem (\ref{eq primal-dual}) into a higher-dimensional HSDS, transforming the task of optimizing (\ref{eq primal-dual}) into finding a non-trivial solution to a set of equations, thereby significantly reducing the computational complexity. In other words, the KKT conditions are homogenized into the following system:
\begin{equation}
    \left\{ \begin{aligned}
        & \bm{A}\bm{x} + \bm{s} - \bm{b}\tau = \bm0, \ \text{(Primal feasibility residual)}, \\
        & \bm{A}^\top \bm{y} + \bm{c}\tau = \bm0, \ \text{(Dual feasibility residual)}, \\
        & -\bm{b}^\top\bm{y} - \bm{c}^\top \bm{x} -\kappa = 0, \ \text{(Duality gap residual)}, \\
        & \bm{s}^\top \bm{y}+ \tau\kappa= \bm0, \ \text{(Complementary slackness)},   \\
        & \bm{s} \in \mathcal{K}, \ \bm{y} \in \mathcal{K}^*, \tau \geq 0, \kappa \geq 0.
    \end{aligned} \right. \label{eq HSDS}
\end{equation}
where the decision variables could be collected as $\bm h=(\bm x, \bm y, \bm s, \tau, \kappa)\in\mathbb{R}^{D_2}$. $\tau$ is a scaling factor. $\kappa$ is an optimality indicator.

Subsequently, (\ref{eq HSDS}) can be compactly rewritten as a linear system given the auxiliary matrix $\Omega$:
\begin{equation}
    F(\bm h, \Omega) = \Omega \bm h + \bm t = \bm 0  \label{eq linear}
\end{equation}
where $ \bm t\in \mathbb{R}^{D_1}$ is the slack variable.

The non-trivial solution $\bm h^*\ne0$ of (\ref{eq linear}) is actually an implicit function of $\Omega$, written as $\bm h^*=g(\Omega)$. The derivative of $\bm h^*$ with respect to $\Omega$ can be calculated as follows without acquiring the specific mathematical form of $g(\cdot)$ by utilizing the implicit function theorem \cite{krantz2002implicit}:
\begin{equation}
    \nabla_g(\Omega)=-[(\partial F/\partial \bm h)^{-1}](\partial F/\partial \Omega). \label{eq diff HSDS}
\end{equation}

\subsubsection{Solution Mapping}
The HSDS in (\ref{eq HSDS}) is always feasible, as it is guaranteed to have a trivial solution, i.e., $\bm 0$. If we could find a non-trivial solution satisfying $\tau > 0$, all the decision variables $\bm{h}$ can be scaled by dividing by $\tau$. Due to the complementary slackness conditions, we have $\kappa=0$. The resulting optimal solution $(\bm x^{\diamond}/\tau, \bm y^{\diamond}/\tau,\bm s^{\diamond}/\tau)$ then precisely satisfies the KKT conditions in (\ref{eq KKT}), demonstrating that this solution is exactly the optimal solution to the problem (\ref{eq primal-dual}). Therefore, given the optimal solution of HSDS $\bm h^*$, we have:
\begin{equation}
    (\bm{x}^*,\bm{y}^*,\bm{s}^*)=(\frac{\bm x^{\diamond}}{\tau}, \frac{\bm y^{\diamond}}{\tau},\frac{\bm s^{\diamond}}{\tau})=\phi(\bm x^{\diamond}, \bm y^{\diamond}, \bm s^{\diamond}, \tau, \kappa).
\end{equation}
\begin{equation}
    \nabla_{\phi}(h)=\begin{bmatrix}
        \bm I_p/\tau & \bm 0 & \bm 0 & -\bm x/\tau^2 & \bm 0 \\
        \bm 0 & \bm I_d/\tau & \bm 0 & -\bm y/\tau^2 & \bm 0 \\
        \bm 0 & \bm 0 & \bm I_d/\tau & -\bm s/\tau^2 & \bm 0 
    \end{bmatrix}
\end{equation}
where $\bm I_p \in \mathbb{R}^p$ and $\bm I_d \in \mathbb{R}^d$ are identity matrices.

\begin{algorithm} 
    \caption{The Proposed Optimization-embedded E2E Learning Algorithm for Power System RSE}
    \label{ag.workflow}  
    \begin{algorithmic}[1]  
        \State \textbf{Dataset preparation:} Prepare the full dataset $\mathcal{D}=\{\bm{z},\bm{V} \}$, containing measurement data $\bm{z}$ as input feature, state variables $\bm{V}$ as output labels. Hyperparameters (learning rate $\alpha$, weight decay factor $\zeta$, batch size $B$, training epoch $E$, physical loss weight factor $\rho$). 
        \State \textbf{Dataset separation:} Dataset $\mathcal{D}$ is separated into a training dataset $\mathcal{D}_{\text{train}}$ and a testing dataset $\mathcal{D}_{\text{test}}$;
        \State \textbf{Layer construction:} Formulate parametrized relaxed RSE problem based on $(\mathrm{P3})$ and construct the $\mathrm{Opt-Layer}$;
        \State \textbf{Offline training:}
        \State Input training dataset $\mathcal{D}_{\text{train}}$, initial measurement weights $\bm{w}_m^0$;
        \For{epoch $=1,...,E$}
            \State \textbf{Forward propagation:}
            \State \textit{Decision-making}: solve the parametrized relaxed  
            \Statex \hspace{\algorithmicindent} $\quad$ problems in each batch via $\mathrm{Opt-Layer}$;
            \State \textit{Post-processing}: map the solutions of the relaxed
            \Statex \hspace{\algorithmicindent} $\quad$ problems for recovery, obtaining estimated $\hat{\bm{V}}$;
            \State \textbf{Backward propagation:} 
            \State Calculate the derivative of surrogate loss function 
            \Statex \hspace{\algorithmicindent}  with respect to all the learnable parameters $\bm{\mu}$ as $\nabla_{\mu}\hat{\mathcal{L}}$;
            \State \textbf{Parameters updating:}
            \State Update the learnable parameters of NN.
        \EndFor
        \State Obtain the trained E2E learning model.
        \State \textbf{Online application:}
        \State \textit{Decision-making:} Input testing dataset $\mathcal{D}_{\text{test}}$, output estimated state variables $\hat{\bm{V}}$.
        \State \textit{Performance evaluation:} Calculate accuracy loss $\mathcal{L}^{\text{acc}}$, measurement residuals $\mathcal{L}^{\text{huber}}$, and regularization term $\mathcal{L}^{\text{reg}}$.
    \end{algorithmic}
\end{algorithm}

\subsection{Training and Inference Process}
Building upon the NN architecture established in Section \ref{sec2} and the relaxed SOCP-based RSE problem $\mathrm{(P3)}$ derived in Section \ref{sec3}, Algorithm 1 details the proposed optimization-embedded E2E learning algorithm for power system RSE. The procedure commences with the generation of a dataset. The input features consist of measurements $\bm{z}$, while the output labels are the corresponding system states $\bm{V}$. In real-world physical power systems, acquiring absolute ground-truth states is a well-known challenge. To address this limitation in practice, the training dataset can be constructed through two primary approaches. First, system operators can utilize vast amounts of historical load and generation profiles to run AC power flow calculations, thereby generating a massive dataset of physically consistent states (serving as labels) and their corresponding ideal measurements. Noise and extreme outliers are then artificially injected to create the input features, which is the approach adopted in our case studies. Alternatively, when relying purely on historical SCADA data, operators can employ computationally intensive offline bad data processing and manual verification to generate highly accurate "pseudo-labels". Since offline training is not bound by strict real-time computational constraints, both approaches ensure the availability of high-quality training data, fully supporting the "offline training, online deployment" paradigm. After defining necessary hyperparameters and partitioning the dataset $\mathcal{D}$ into training set $\mathcal{D}_{\text{train}}$ and testing set $\mathcal{D}_{\text{test}}$, the relaxed problem $(\mathrm{P3})$ is parametrically modeled to construct the differentiable $\mathrm{Opt-Layer}$ \cite{agrawal2019differentiable}.

During the training stage, the NN processes batches from $\mathcal{D}_{\text{train}}$. The set of learnable parameters for the entire NN $\bm{\mu}$ includes the measurement weights $\bm{w}_m$ within the $\mathrm{Opt-Layer}$, as well as the weights and biases of the post-processing layers. In the FP, as described in Section \ref{sec2}, the $\mathrm{Opt-Layer}$ first solves the relaxed problems $(\mathrm{P3})$ in the batch, and the post-processing layers then map these solutions back to the original variable space. In the BP, derivatives are computed via the chain rule. Specifically, the derivatives of the loss function $\hat{\mathcal{L}}$ with respect to the parameters of post-processing layers are calculated first, followed by the derivatives with respect to the learnable measurement weights $\bm{w}_m$ using the method detailed in Section IV-B. Subsequently, all learnable parameters $\bm{\mu}$ across the NN are updated. Moreover, to ensure efficient convergence and reduce sensitivity to hyperparameter tuning, the adaptive moment estimation (Adam) optimizer is employed. Furthermore, the weight decay technique is incorporated to mitigate the risk of overfitting.

Once training is complete, the resulting NN provides a direct mapping from raw measurement inputs to the estimated system states. This enables online deployment, a significant advantage over traditional optimization-based approaches, whose computational demands often limit their potential for real-time applications. Detailed runtime and computational complexity analysis are presented in Section \ref{sec V-comp}.

\subsection{Sparsification}
A critical consideration for practical implementation is the computational burden associated with the $\mathrm{Opt-Layer}$, particularly for large-scale systems. Although the dimension of the final estimated states is $2N$, the decision variables of the relaxed problem $(\mathrm{P3})$ include $\mathrm{Re}(X_{ij})\in\mathbb{R}^{N\times N}$ and $\mathrm{Im}(X_{ij})\in\mathbb{R}^{N\times N}$. This results in a stacked $\mathrm{Opt-Layer}$ output dimension of $N + 2N^2$ (i.e., $\mathcal{O}(N^2)$), indicating a heavy computational cost during FP. 

The inherent sparsity of the nodal admittance matrix $Y$ is utilized to address this challenge. In a power system, $Y_{ij}=0$ signifies that no direct branch connects bus $i$ and $j$. Consequently, the corresponding element $X_{ij}$ is also zero. Therefore, in our work, only $\mathrm{Re}(X_{ij})$ and $\mathrm{Im}(X_{ij})$ are treated as active decision variables for the bus pairs $(i, j)$ where $Y_{ij} \neq 0$. All the other elements of $X_{ij}$ are fixed to zero. This strategy drastically reduces the number of decision variables and constraints, realizing significantly accelerated training.

To quantify the computational improvement achieved by this sparsification, we could consider the complexity of solving the SOCP problem using standard interior point methods (IPMs). The dominant computational burden in IPMs lies in solving the $\mathcal{O}(N) \times \mathcal{O}(N)$ Newton linear system at each iteration, which requires $\mathcal{O}(N^3)$ operations in the worst case. Without sparsification, the complexity of solving $(\mathrm{P3})$ would be $\mathcal{O}((N^2)^3)=\mathcal{O}(N^6)$, which would be intractable for large-scale power systems. By sparsification, the number of decision variables drastically drops to $\mathcal{O}(N)$. This reduction brings the theoretical worst-case complexity per IPM iteration down to $\mathcal{O}(N^3)$. In practical implementations, modern SOCP solvers would exploit the sparsity of the power system and advanced sparse matrix factorization techniques to further reduce the complexity.

\section{Case Studies}\label{sec5}
\subsection {Parameter setup}\label{sec V-setup}
To validate the performance of the proposed optimization-embedded E2E learning framework, extensive simulations are conducted on 8 testing power systems, encompassing both radial (Baran 33-bus and Mantovani 136-bus cases) and meshed topologies (WSCC 9-bus, IEEE 14-bus, New England 39-bus, IEEE 57-bus, IEEE 300-bus, and PEGASE 1354-bus cases). The grid parameters of the above test systems are extracted from the MATPOWER 7.1 case files. Unless otherwise specified in the subsequent subsections, the measurement dataset is assumed complete, including all the nodal voltage magnitudes, branch power flows, and nodal power injections. The estimation performance under partial measurement observability is presented in Section \ref{sec V-observe}.

The dataset for each system is constructed as follows. First, true system states are obtained by running probabilistic power flow calculations using the given parameters in MATPOWER case files. Specifically, Gaussian noise with a standard deviation of 0.02 p.u. is applied to all loads to perturb the system, generating a total of 2000 samples. Measurements are then simulated by firstly adding a basic zero-mean Gaussian noise, with a standard deviation of $\sigma_m = 0.001$ p.u., to these ground-truth values. Considering potential outliers in the measurement dataset, a proportion $\eta$ of the measurements are selected to be injected with scaled zero-mean Gaussian noise in addition, where the standard deviation is amplified by a scaling factor $k$. Unless otherwise specified in the subsequent subsections, the outlier penetration rate is set to $\eta=15\%$ and the scaling factor is $k=30$. The estimation performance under different noise contamination levels and structured non-Guassian noise is presented in Section \ref{sec V-robustness}. For simplicity, the coefficient $\rho_r$ in the objectives of $\mathrm{(P2)}$ and $\mathrm{(P3)}$ is set as 1. To ensure smoothness and differentiability, the Huber loss is employed as an alternative to the exact L1-norm, with the parameters $\delta$ and $\epsilon_p$ are together set as 1e-5. Given that the standard deviation of the basic measurement noise in our dataset is $\sigma_m= 0.001$ p.u., choosing $\delta = 1e-5$ (a value significantly smaller than the noise level) ensures that the Huber loss tightly approximates the exact L1-norm while providing the necessary differentiability for the learning framework.

The hyperparameters configured for the E2E learning framework are set as follows: learning rate $\alpha= 1e-3$, weight decay factor $\zeta=5e-4$, batch size $B=32$, training epoch $E=100$. In particular, a sensitivity analysis on the selection of the physical loss weight $\rho$ within the hybrid loss function is presented in Section \ref{sec V-weight}. The exponential decay rates of Adam, denoted by $\beta_1$ and $\beta_2$, are set as $0.9$ and $0.999$ by default. For each testing system, the 2000 samples are partitioned into a training set $\mathcal{D}_{\text{train}}$ and a testing set $\mathcal{D}_{\text{test}}$ at an 8:2 ratio. All simulations are implemented by Matlab 2024a and Python 3.12 with PyTorch 2.8.0 and CVXPY packages on a personal computer and optimization problems are solved by Gurobi. The hardware environment is Intel(R) Xeon(R) Gold 6430 CPU with 120GB of RAM and NVIDIA GeForce RTX 4090 GPU on workstations.

\subsection{Robustness Comparison Using Different Estimators}\label{sec V-estimator}
To demonstrate the robustness of the WLAV estimator against varying levels of outlier penetration, simulations on the WSCC 9-bus testing system are conducted. Five sets of parameters are set as follows. Set 1: $\eta=0\%$, without outliers. Set 2: $\eta=15\%, k=10$. Set 3: $\eta=15\%, k=50$. Set 4: $\eta=30\%, k=10$. Set 5: $\eta=30\%, k=50$. Under each setting, 200 scenarios are generated.

\begin{figure}
    \centering
    \includegraphics[width=\linewidth]{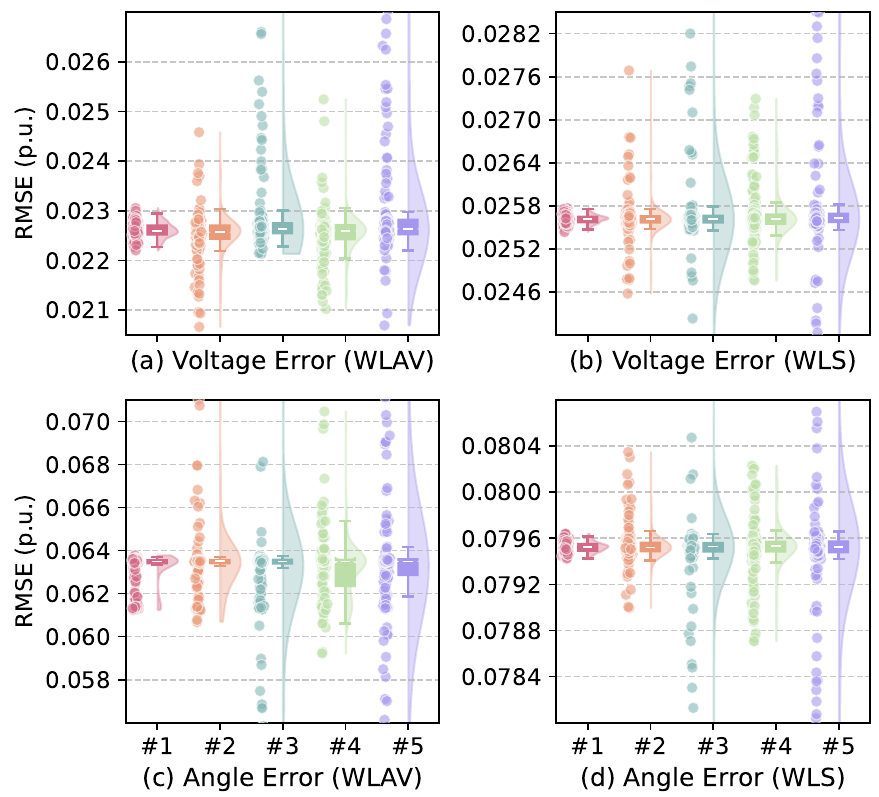}
    \caption{Estimation Robustness Comparison on WSCC-9 Bus System.}
    \label{fig:Case9}
\end{figure}

Fig. \ref{fig:Case9} compares the root mean square error (RMSE) of estimated voltage magnitude and phase angle using WLAV-based and WLS-based estimators. In Fig. \ref{fig:Case9} (a), the medians of voltage magnitude RMSE using WLAV-based estimator are between [0.022, 0.023] p.u., while the results of WLS-based estimator are within the range of [0.0254, 0.0258] p.u.. Similarly, the medians of angle RMSE using WLAV-based and WLS-based estimators fall into the range of [0.063, 0.064] p.u. and [0.0794, 0.0796] p.u., respectively. Obviously, the errors obtained from the WLAV-based estimator are all statistically significantly lower under different settings of outlier penetration. Also, as $\eta$ and $k$ increase, the distribution of errors becomes more dispersed, reflecting a higher proportion of large estimation errors. These results confirm that the WLS-based estimator is more sensitive to outliers due to the quadratic residual term in its objective, whereas the WLAV-based estimator demonstrates higher robustness.

\begin{table*}
\footnotesize
\centering
\caption{Estimation Performance Comparison across different learning and optimization models.}
\label{tab.models}
\begin{tabularx}{0.95\textwidth}{c|ccc|ccc|ccc|ccc}
\toprule
\specialrule{0em}{0.3pt}{0.5pt}
\toprule
\multirow{2}{*}{\centering \textbf{Model}} & $\mathcal{L}^{\text{acc}}$ & $\mathcal{L}^{\text{huber}}$ & $\mathcal{L}^{\text{reg}}$ & $\mathcal{L}^{\text{acc}}$ & $\mathcal{L}^{\text{huber}}$ & $\mathcal{L}^{\text{reg}}$ & $\mathcal{L}^{\text{acc}}$ & $\mathcal{L}^{\text{huber}}$ & $\mathcal{L}^{\text{reg}}$ & $\mathcal{L}^{\text{acc}}$ & $\mathcal{L}^{\text{huber}}$ & $\mathcal{L}^{\text{reg}}$ \\
\cmidrule(r){2-4} \cmidrule(l){5-7} \cmidrule(l){8-10} \cmidrule(r){11-13}
& \multicolumn{3}{c|}{\textbf{WSCC-9}} & \multicolumn{3}{c|}{\textbf{IEEE-14}} & \multicolumn{3}{c}{\textbf{Baran 33}} & \multicolumn{3}{c}{\textbf{New England 39}}  \\
\midrule
M1 & 0.00086 &	\underline{2e-6}  &	\underline{0.00004} &	0.00117 &	\underline{1e-6}	&0.00009 &	0.00233 &	\underline{0.00013} 	& 0.00705 	& 0.00338 	& \underline{0.00003} 	& \underline{0.00051} \\
M2 & 0.00157 	&0.09892 &	0.00170 &	0.00152 &	0.03235 &	0.00209 &	0.00070 &	0.09943 &	0.02501 &	0.00251 &	0.74316 &	0.05581  \\
M3 & 0.00158 &	0.09881 &	0.00167 &	0.00151 &	0.03235 &	0.00210 &	0.00574 &	0.11153 &	0.02824 &	0.02092 &	0.77326 &	0.04930  \\
M4 & 0.00195 &	0.10032 &	0.00077 &	0.00335 &	0.04110 &	\underline{1.1e-6}	& 0.00051 &	0.14578 &	\underline{0.00222} &	0.00272 &	0.75936 &	0.04404 \\
M5 & 0.00007 &	0.16132 &	0.04366 &	0.00003 &	0.04902 &	0.12990 &	0.00002 &	0.05218 &	0.21416 &	\underline{0.00028} &	0.86295 &	0.39267  \\
M6 & \underline{0.00002} &	0.17031 &	0.04858 &	\underline{3e-6} &	0.04996 &	0.13296 &	0.00001 &	0.04940 &	0.21013 &	0.00396 &	0.82933 &	0.31141 \\
M7 & 0.00148 &	0.16533 &	0.02330 &	0.00302 &	0.04590 &	0.00343 &	0.00015 &	0.14170 &	0.05990 &	0.00092 &	1.97565 &	0.99436 \\
M8 & 0.00024 &	0.15320 &	0.04097 &	0.00023 &	0.04498 &	0.10727 &	\underline{1e-6} & 0.04546 &	0.20926 &	0.00081 &	0.75453 &	0.29484 \\
M9 & 0.00884 &	0.13850 &	0.02647 &	0.00090 &	0.04150 &	0.08884 &	3e-6 & 0.04546 &	0.18893 &	0.00794 &	0.84087 &	0.29150 \\
\midrule
\textbf{Model} & \multicolumn{3}{c|}{\textbf{IEEE-57}} & \multicolumn{3}{c|}{\textbf{Mantovani 136}} & \multicolumn{3}{c}{\textbf{IEEE-300}} & \multicolumn{3}{c}{\textbf{PEGASE 1354}}  \\
\midrule
M1 & 0.00077 &	\underline{3e-6} &	\underline{0.00003} &	0.11526 & \underline{0.00682} & 0.00448 & 0.12615 & \underline{0.00353} & \underline{0.00588}  & 	0.19293 &	\underline{0.27864} &	0.09449 \\
M2 & 0.00163 &	0.03697 &	0.00281 &	0.08390 & 0.08750 & 0.01339 & 0.04873 & 0.42772 & 0.02544  &	0.58522 &	0.58723 &	0.14185 \\
M3 & 0.00417 &	0.03864 &	0.00264 &	0.26659 & 0.08242 & 0.00600 & 0.10709 & 0.46061 & 0.03192  &	0.58359 &	0.58724 &	0.12585 \\
M4 & 0.00248 &	0.04136 &	0.00291 &	0.01039 & 0.00915 & \underline{0.00028} & 0.22904 & 0.48745 & 0.00732  &	0.55622 &	0.58726 &	\underline{0.00001} \\
M5 & \underline{0.00001} &	0.04673 &	0.28103 &	0.00001 &	0.48223 &	0.31307 &	0.00357 &	0.48823 & 	3.34119 &	0.00845 &	0.62050 &	11.374 \\
M6 & 0.00010 &	0.04825 &	0.29440 &	0.00001 &	0.27840 &	0.18514 &	0.00204 &	0.54965 &	3.84879 &	\underline{0.00222} &	0.79161 &	14.669 \\
M7 & 0.00107 &	0.07163 &	0.10961 &	0.00029 &	0.00908 &	0.00046 &	0.02020 &	2.11877 &	14.6565 & 0.00914 &	9.14470 &	269.80 \\
M8 & 0.00139 &	0.03135 &	0.14597 &	\underline{1e-6} &	0.00276 &	0.03173 &	\underline{0.00138} &	0.32248 &	3.29102 &	0.03240 &	0.43536 &	15.804 \\
M9 & 0.01041 &	0.03839 &	0.10219 &	2e-6 &	0.00276 &	0.02966 &	0.02280 &	0.47838 &	3.60314 &	0.16684 &	0.57658 &	17.923 \\
\bottomrule
\specialrule{0em}{0.5pt}{0.3pt}
\bottomrule
\end{tabularx}
\end{table*}

\subsection{Estimation Performance Comparison Across Different Learning and Optimization Models}\label{sec V-models}
\subsubsection{Simulation settings}
To validate the RSE performances obtained by the proposed E2E learning framework, comparative experiments using seven learning models and two optimization-based models are carried out. Specifically, the proposed framework is denoted as Model 1 (M1), whose NN structure is the combination of a differentiable $\mathrm{Opt-Layer}$ and a set of fully connected post-processing layers, as described in Section \ref{sec2}. The $\mathrm{Opt-Layer}$ is constructed from the relaxed RSE problem itself, ensuring strict optimality conditions embedding during BP. As illustrated in Algorithm 1, for our model M1, the input context of the NN includes the measurement data and the initial value for the learnable measurement weights $\bm{w}_m^0$, which is set to $\bm{1}$ to ensure numerical stability.

To provide a comprehensive evaluation, the other six learning models comprise convolutional neural network (CNN), standard fully connected neural network (FCNN), and graph attention network v2 (GATv2) \cite{brody2022how}. Models 2, 3, and 4 (M2, M3, and M4) correspond to CNN, FCNN, and GATv2 trained with the surrogate hybrid loss function $\hat{\mathcal{L}}$ (i.e., serving as PINNs), where the physical loss weight $\rho$ is set to 1 for simplicity. To highlight the role of the hybrid loss function, Models 5, 6, and 7 (M5, M6, and M7) utilize the same CNN, FCNN, and GATv2 architectures but are trained using a pure MSE loss function. Compared to the standard static GAT, GATv2 allows a node to dynamically adjust its attention weights towards different neighbors based on their specific current states, which is well-suited for the RSE task. Finally, Models 8 and 9 (M8 and M9) represent traditional optimization-based RSE solved by commercial solvers like Gurobi. M8 solves WLAV-based RSE under  the AC power flow formulation via successive linear programming (SLP) \cite{yang2016optimal}. M9 solves the  WLAV-based RSE under the second-order conic relaxation (as modeled in $\mathrm{(P3)}$) \cite{chen2020robust}. Together, these nine models establish a comprehensive comparison across optimization-embedded NN, PINNs, pure data-driven NNs, and optimization methods. To demonstrate the generalizability and scalability of the proposed E2E learning framework, these \textit{nine} models are all validated across the \textit{eight} test systems introduced in Section \ref{sec V-setup}.

\subsubsection{Results analysis}
Table I compares the performances of the nine models on three metrics: MSE-based regression error $\mathcal{L}^{\text{acc}}$, Huber-based measurement residuals $\mathcal{L}^{\text{huber}}$ evaluating the physical consistency, and the regularization term $\mathcal{L}^{\text{reg}}$. Collectively, the proposed M1 outperforms traditional optimization models (M8, M9) in terms of physical consistency, which in turn generally surpass PINNs (M2-M4) and pure data-driven models (M5-M7). The best results among learning models for each metric are highlighted with an underline.

Unlike OPF problems where physical constraints must be strictly satisfied, RSE is fundamentally a robust regression problem. Since input measurements inherently contain noise, it is mathematically almost impossible to find a state that perfectly satisfies all measurement equations with zero error. \textit{Therefore, in the context of RSE, lower measurement residuals indicate that the estimated states are closer to the true physical manifold of the power system (higher physical consistency).}

According to Table I, M1 achieves the lowest measurement residuals ($\mathcal{L}^{\text{huber}}$) across all test systems. The residuals of M1 are often several orders of magnitude lower than those of PINNs (e.g., $2\mathrm{e}{-6}$ vs. 0.09892 in WSCC-9, and $3\mathrm{e}{-6}$ vs. 0.03697 in IEEE-57). This superiority stems from the $\mathrm{Opt-Layer}$, which explicitly solves the relaxed RSE problem during the FP and rigorously embeds the KKT conditions during BP.

Regarding the other learning models, pure data-driven models (M5-M7) achieve low MSE but exhibit substantially higher physical losses, illustrating that a statistically accurate solution is not necessarily a physically consistent one. PINNs (M2-M4) mitigate this by incorporating soft penalties on the inconsistency. Notably, the GATv2-based models (M4 and M7) perform comparably to FCNN/CNN-based models on small-scale systems, but demonstrate remarkable advantages on large-scale systems. For instance, in the Mantovani 136-bus system, M4 achieves a significantly lower residual (0.00915) and regularization term (0.00028) compared to M2 and M3. This is attributed to the dynamic attention mechanism of GATv2, which can adaptively assign lower attention scores to neighbors corrupted by severe outliers, thereby enhancing the robustness of spatial feature extraction in complex topologies.

Comparing the traditional optimization models (M8 and M9) provides further insights. For radial systems (e.g., Baran 33 and Mantovani 136), M8 and M9 achieve distinguished estimation accuracy. Besides, M8 and M9 yield nearly identical measurement residuals, verifying that the SOCP relaxation is strictly tight for radial distribution networks. For meshed grids, the estimation accuracy of M8 and M9 is comparable with pure MSE-based data-driven methods (M5-M7). M9 performs slightly worse than M8 due to the inherent relaxation gap when the system scale grows. More importantly, both M8 and M9 are outperformed by the proposed M1 in terms of physical consistency. The underlying reason is that traditional optimization solvers are "myopic": they solve equations sample-by-sample from scratch, fitting values without prior knowledge of the noise distribution. This contributes to the distinguished performance of solvers in terms of accuracy. In contrast, the optimization-embedded data-driven learning framework M1 possesses a "global" perspective. By observing massive samples during offline training, the NN implicitly encodes the statistical distribution of noises and outliers into its weights, empowering the $\mathrm{Opt-Layer}$ to effectively filter out noises and resist outliers during online inference, and finally leads to more physically consistent state estimations.

\begin{table*}
\centering
\footnotesize
\caption{Estimation Performance Comparison under different weights.}
\label{tab.weights}
\begin{tabularx}{\textwidth}{c|c|ccc|ccc|ccc|ccc}
\toprule
\specialrule{0em}{0.3pt}{0.5pt}
\toprule
\multirow{2}{*}{\centering $\rho$}  & \multirow{2}{*}{\centering \textbf{Model}} & $\mathcal{L}^{\text{acc}}$ & $\mathcal{L}^{\text{huber}}$ & $\mathcal{L}^{\text{reg}}$ & $\mathcal{L}^{\text{acc}}$ & $\mathcal{L}^{\text{huber}}$ & $\mathcal{L}^{\text{reg}}$ & $\mathcal{L}^{\text{acc}}$ & $\mathcal{L}^{\text{huber}}$ & $\mathcal{L}^{\text{reg}}$ & $\mathcal{L}^{\text{acc}}$ & $\mathcal{L}^{\text{huber}}$ & $\mathcal{L}^{\text{reg}}$ \\
\cmidrule(r){3-5} \cmidrule(l){6-8} \cmidrule(l){9-11} \cmidrule(r){12-14}
& & \multicolumn{3}{c|}{\textbf{WSCC-9}} & \multicolumn{3}{c|}{\textbf{New England 39}} & \multicolumn{3}{c}{\textbf{IEEE-57}} & \multicolumn{3}{c}{\textbf{IEEE-300}}  \\
\midrule
\multirow{4}{*}{\centering 1} & M1 & \underline{0.00086} & 	\underline{2e-6} &	\underline{0.00004}  &	0.00338  &	\underline{0.00003} &	\underline{0.00051}  &	\underline{0.00077}  &	\underline{3e-6} &	0.00003  &	0.12615 	 & \underline{0.03536}	 & 0.05880  \\
& M2 & 0.00157  &	0.09892  &	0.00170  &	\underline{0.00251}  &	0.74316  &	0.05581  &	0.00163  &	0.03697  &	0.00281  &	\underline{0.04873} &	0.42772 &	0.02544   \\
& M3 & 0.00158  &	0.09881  &	0.00167  &	0.02092  &	0.77326  &	0.04930  &	0.00417  &	0.03864  &	0.00264  &	0.10709 &	0.46061 &	0.03192   \\
& M4 & 0.00195  &	0.10032  &	0.00077  &	0.00272  &	0.75936  &	0.04404  &	0.00248  &	0.04136  &	0.00291  &	0.22904 &	0.48745 &	\underline{0.00732}   \\
\midrule
\multirow{4}{*}{\centering 0.1} & M1 & \underline{0.00061}  &	\underline{0.00087}  &	0.00164  &	0.00264  &	\underline{0.00793}  &	\underline{0.00981}  &	\underline{0.00073}  &	\underline{0.00273}  &	0.00115  &	\underline{0.00073}  &	\underline{0.00273}  &	\underline{0.01149}  \\
& M2 & 0.00108  &	0.09907  &	0.00398  &	\underline{0.00224}  &	0.74373  &	0.05639  &	0.00125  &	0.03646  &	0.00517  &	0.01934 &	0.43774 &	0.10432  \\
& M3 & 0.00111  &	0.09895  &	0.00391  &	0.01016  &	0.78209  &	0.04017  &	0.00114  &	0.03651  &	0.00512  & 0.02230 &	0.44036 &	0.09032  \\
& M4 & 0.00178  &	0.10061  &	\underline{0.00129}  &	0.06103  &	0.90781  &	0.01265  &	0.01508  &	0.05927  &	\underline{0.00038}  &	0.01508 &	0.48291 &	0.02762  \\
\midrule
\multirow{4}{*}{\centering 10} & M1 & \underline{0.00154}  &	\underline{0.00056}  &	\underline{0.00003}  &	\underline{0.00441}  &	\underline{0.03145}  &	\underline{0.00328}  &	\underline{0.00149}  &	\underline{0.00105}  &	\underline{0.00005}  &	0.47860 &	\underline{0.00370} &	0.00236   \\
& M2 & 0.00207  &	0.09879  &	0.00138  &	0.01246  &	0.76047  &	0.04759  &	0.00351  &	0.03790  &	0.00239  &	0.49166 &	0.50380 &	\underline{0.00081}  \\
& M3 & 0.00606  &	0.10071  &	 0.00134  &	0.02428  &	0.79941  &	0.03247  &	0.00633  &	0.03975  &	0.00228  &	0.51308 &	0.50387 &	0.00210  \\
& M4 & 0.00433  &	0.10026  &	0.00094  &	0.16786  &	0.90152  &	0.01687  &	0.03561  &	0.06105  &	0.00149  &	\underline{0.15001} &	0.48327 &	0.02471  \\
\bottomrule
\specialrule{0em}{0.5pt}{0.3pt}
\bottomrule
\end{tabularx}
\end{table*}

\subsection{Estimation Performance Comparison Across Different Physical Loss Weight Selection}\label{sec V-weight}
\subsubsection{Simulation settings}
To systematically address the hyperparameter sensitivity, we evaluated the estimation performances under different selections of physical loss weight $\rho$ across four representative system scales (WSCC-9, New England 39, IEEE-57, and IEEE-300). As presented in Table II, these four systems are selected to comprehensively reveal the scaling trends without redundant computational overhead. We compared the proposed M1 against all the PINN-based models (M2-M4) under three distinct scenarios: an accuracy-focused approach with $\rho=0.1$, a balanced approach with $\rho=1$, and a physics-focused approach with $\rho=10$.

\subsubsection{Results analysis}
Table II demonstrates the difference in how the hyperparameter $\rho$ affects the proposed optimization-embedded framework versus PINNs. The fundamental trade-off is evident across all models and system sizes: as $\rho$ increases, the MSE-based regression error ($\mathcal{L}^{\text{acc}}$) inevitably increases. However, the trends in physical consistency ($\mathcal{L}^{\text{huber}}$) diverge. For the PINN-based models (M2-M4), increasing $\rho$ from $0.1$ to $10$ fails to yield a corresponding reduction in measurement residuals. Across all four systems, their $\mathcal{L}^{\text{huber}}$ remains stubbornly high (e.g., hovering around 0.74-0.79 for the 39-bus system and 0.42-0.51 for the 300-bus system) regardless of the $\rho$ value. This indicates that in high-dimensional, non-convex power systems, relying purely on soft penalties in the loss function may hit a bottleneck.

In contrast, M1 consistently maintains lower measurement residuals than those of M2-M4 across all $\rho$ settings and system scales. Remarkably, even in the accuracy-focused scenario ($\rho=0.1$), M1 achieves exceptional physical consistency (e.g., $\mathcal{L}^{\text{huber}}$=0.00273 for the IEEE-300 system, compared to 0.43774 for M2). This proves that the superior physical consistency of M1 is inherently guaranteed by the structural embedding of the KKT conditions within the $\mathrm{Opt-Layer}$, rather than relying heavily on the external loss weight $\rho$.

Based on these systematic observations across different system scales, we provide the following \textit{hyperparameter tuning suggestions for practitioners}. When deploying PINNs (M2-M4), selecting $\rho$ presents a dilemma, as a large $\rho$ heavily degrades statistical accuracy while offering marginal gains in physical consistency. Conversely, when deploying our proposed framework (M1), practitioners are relieved from this tuning burden. The $\mathrm{Opt-Layer}$ structurally enforces the physical consistency during the BP. Practitioners are recommended to select a relatively small or moderate weight (e.g., $\rho \in [0.1, 1]$) when deploying M1. As demonstrated in Table II, this setting allows the NN to focus primarily on extracting features to minimize the regression error ($\mathcal{L}^{\text{acc}}$), while the embedded $\mathrm{Opt-Layer}$ naturally and robustly handles the physical consistency, yielding superior overall estimation performance.

\subsection{Performance Comparison Under Partial Observability}\label{sec V-observe}
\subsubsection{Simulation settings}
To rigorously address practical concerns regarding incomplete measurement sets, we systematically evaluate the proposed framework under varying degrees of observability. We select M2 as the baseline, as it demonstrated relatively more stable performance among the PINN-based models. Three distinct scenarios are designed:
\begin{itemize}
    \item \textit{Case 1 (Ideal Observability):} 100\% completed measurement availability, serving as the ideal baseline.
    \item \textit{Case 2 (Measurement observability with zero redundancy):} The system only has access to all nodal voltage magnitudes and the active power flows of branches that exactly form a spanning tree. This satisfies the conditions presented in Appendix A, but with minimal redundancy.
    \item \textit{Case 3 (Measurement observability violating the conditions in Appendix A):} built on Case 2, 20\% of nodal voltage measurements are randomly dropped, and only 70\% of the branches in the spanning tree are retained, violating the conditions.
\end{itemize}

To quantitatively analyze the estimation performances and relaxation tightness, we also track the $\mathcal{L}^{\text{acc}}$, $\mathcal{L}^{\text{huber}}$, $\mathcal{L}^{\text{reg}}$, the maximum phase angle difference $|\Delta \theta|$ across all branches, and the average eigenvalue ratio $\bar{\lambda}$ (defined in Section III-D) for all the eight test systems, as comprehensively presented in Table III.

\begin{table*}
\centering
\small
\caption{Performance comparison under partial observability.}
\label{tab.observe}
\begin{tabularx}{0.85\textwidth}{c|c|c|c|ccc|ccc}
\toprule
\specialrule{0em}{0.3pt}{0.5pt}
\toprule
\multirow{2}{*}{\centering Test system} & \multirow{2}{*}{\centering $|\Delta \theta|$} & \multirow{2}{*}{\centering Case} & \multicolumn{4}{c|}{\textbf{M1}} & \multicolumn{3}{c}{\textbf{M2}} \\
\cmidrule(r){4-7} \cmidrule(r){8-10} 
 & & & $\bar{\lambda}$ & $\mathcal{L}^{\text{acc}}$ & $\mathcal{L}^{\text{huber}}$ & $\mathcal{L}^{\text{reg}}$ & $\mathcal{L}^{\text{acc}}$ & $\mathcal{L}^{\text{huber}}$ & $\mathcal{L}^{\text{reg}}$ \\
\midrule
\multirow{3}{*}{\centering \textbf{WSCC-9}} & \multirow{3}{*}{\centering 8.17$\degree$} & Case 1 & 1  & 0.00086  & 2e-6 & 0.00004  & 0.00157  & 0.09892  & 0.00170  \\
 &  & Case 2 & 0.9940  & 0.00120  & 0.00034  & 0.00006  & 0.00052  & 0.01061  & 0.01194  \\
 &  & Case 3 & 0.9956  & 0.00099  & 0.00027  & 0.00004  & 0.00072  & 0.00578  & 0.00784  \\
\midrule
\multirow{3}{*}{\centering \textbf{IEEE-14}} & \multirow{3}{*}{\centering 9.21$\degree$} &
Case 1 & 1  & 0.00117  & 1e-6 & 0.00009  & 0.00152  & 0.03235  & 0.00209 \\
 &  & Case 2 & 0.9980  & 0.00127  & 0.00055  & 0.00034  & 0.00077  & 0.02567  & 0.01762 \\
 &  & Case 3 & 0.9988  & 0.00117  & 0.00060  & 0.00006  & 0.00083  & 0.02226  & 0.01991 \\
\midrule
\multirow{3}{*}{\centering \textbf{Baran 33}} & \multirow{3}{*}{\centering 0.24$\degree$} &
Case 1 & 1  & 0.00233  & 0.00013  & 0.00705  & 0.00070  & 0.09943  & 0.02501 \\
 &  & Case 2 & 0.9999  & 0.00246  & 0.03173  & 0.06728  & 0.00546  & 0.08391  & 0.06909 \\
 &  & Case 3 & 0.9999  & 0.02854  & 0.07040  & 0.56716  & 0.00514  & 0.06891  & 0.06258 \\
\midrule
\multirow{3}{*}{\centering \textbf{New England 39}} & \multirow{3}{*}{\centering 12.53$\degree$} &
Case 1 & 0.9999  & 0.00338  & 0.00003  & 0.00051  & 0.00251  & 0.74316  & 0.05581 \\
 &  & Case 2 & 0.9978  & 0.00314  & 0.01926  & 0.01088  & 0.00687  & 0.06823  & 0.22002 \\
 &  & Case 3 & 0.9978  & 0.00325  & 0.01893  & 0.01720  & 0.00331  & 0.03575  & 0.17101 \\
\midrule
\multirow{3}{*}{\centering \textbf{IEEE-57}} & \multirow{3}{*}{\centering 9.60$\degree$} &
Case 1 & 1  & 0.00077  & 3e-6  & 0.00003  & 0.00163  & 0.03697  & 0.00281 \\
 &  & Case 2 & 0.9980  & 0.00105  & 0.00097  & 0.00028  & 0.00115  & 0.02477  & 0.00561 \\
 &  & Case 3 & 0.9985  & 0.00153  & 0.00219  & 0.00178  & 0.00088  & 0.02172  & 0.00523 \\
\midrule
\multirow{3}{*}{\centering \textbf{Mantovani 136}} & \multirow{3}{*}{\centering 1.28$\degree$} &
Case 1 & 0.9999  & 0.11526  & 0.00682  & 0.00448  & 0.08390  & 0.08750  & 0.01339 \\
 &  & Case 2 & 0.9999  & 0.42565  & 0.02738  & 0.00968  & 0.03033  & 0.08790  & 0.02358 \\
 &  & Case 3 & 0.9999  & 0.37497  & 0.03243  & 0.01747  & 0.01516  & 0.04999  & 0.01921 \\
\midrule
\multirow{3}{*}{\centering \textbf{IEEE-300}} & \multirow{3}{*}{\centering 24.13$\degree$} &
Case 1 & 0.9928  & 0.12615  & 0.00353 & 0.00588  & 0.04873  & 0.42772  & 0.02544 \\
 &  & Case 2 & 0.9923  & 0.24652  & 0.12569  & 0.03415  & 0.03641  & 0.36094  & 0.05815 \\
 &  & Case 3 & 0.9927  & 0.28349  & 0.21443  & 0.05989  & 0.02136  & 0.26502  & 0.04976 \\
\midrule
\multirow{3}{*}{\centering \textbf{PEGASE 1354}} & \multirow{3}{*}{\centering 14.48$\degree$} &
Case 1 & 0.9958  & 0.19293  & 0.27864  & 0.09449  & 0.58522  & 0.58723  & 0.14185 \\
 &  & Case 2 & 0.9948  & 0.37583  & 0.55386  & 0.12582  & 0.03621  & 0.68722  & 0.19480 \\
 &  & Case 3 & 0.9944  & 0.46391  & 0.64690  & 0.24976  & 0.03923  & 0.67163  & 0.18753 \\
\bottomrule
\specialrule{0em}{0.5pt}{0.3pt}
\bottomrule
\end{tabularx}
\end{table*}

\subsubsection{Results analysis}
First, the maximum angle differences for all systems remain well below $90^\circ$, satisfying Condition 2 in Appendix A. Furthermore, it is observed that $\bar{\lambda}$ generally decreases as $|\Delta \theta|$ increases, which aligns with the theoretical properties of conic relaxation of power flow models \cite{low2014convexII}.

The eigenvalue ratio $\bar{\lambda}$ serves as a quantitative statistic for relaxation exactness. For small radial distribution networks (e.g., Baran 33), $\bar{\lambda}$ remains extremely close to 1.0 under Cases 1 and 2, indicating a near-optimal rank-1 solution. Interestingly, for small-scale meshed grids (e.g., WSCC-9, IEEE-14) under complete ideal observability (Case 1), $\bar{\lambda}>0.999$ is also achieved. While strict SOCP exactness is a known theoretical property for radial networks in \textit{noiseless} scenarios, the relaxation in our noisy RSE context remains empirically tight not only for radial networks but also for small meshed grids. This is primarily attributed to two factors: (i), the massive redundancy of a complete measurement dataset heavily constrains the feasible region, squeezing the relaxed solution space tightly around the true physical (rank-1) manifold; (ii), the regularization term inherently penalizes non-rank-1 components. However, for complex large-scale meshed grids (e.g., IEEE-300, PEGASE 1354), $\bar{\lambda}$ gradually deviates from 1.0, indicating that the SOCP relaxation exactness is neither theoretically nor empirically guaranteed. For these large-scale meshed grids, although the exact rank-1 solution is hard to obtain, the solution matrix of relaxed RSE problem still retains a highly dominant principal eigenvalue ($\bar{\lambda}>0.990$), representing a sufficiently close approximation. The estimation performance of optimization-embedded M1 still outperforms PINN-based M2.

Despite the inherent relaxation gaps in complex grids, the proposed M1 demonstrates remarkable physical consistency and robustness. Comparing Case 1 and Case 2, both satisfy the theoretical exactness conditions, yet the $\mathcal{L}^{\text{acc}}$ and residuals ($\mathcal{L}^{\text{huber}}$) of M1 naturally increase in Case 2. This is because Case 1 possesses massive measurement redundancy, allowing the $\mathrm{Opt-Layer}$ to effectively denoise. In contrast, Case 2 possesses zero redundancy. With barely enough equations to solve the system, the estimator is forced to absorb the measurement noise entirely.

For Case 3, traditional optimization-based SE solvers typically fail entirely (e.g., due to Jacobian singularity) under such unobservable conditions. Remarkably, despite the violation of exactness conditions, M1 does not suffer a catastrophic failure in Case 3. Instead, it exhibits a \textit{graceful degradation}, maintaining an acceptable level of physical consistency that is still superior to M2. This is attributed to the powerful data-driven prior of the NN. By learning the underlying physical manifold from historical data, the NN could infer the missing measurements, guiding the embedded $\mathrm{Opt-Layer}$ towards a reasonable region.

In contrast, for PINN-based M2, lower observability (Case 3) sometimes yields slightly lower physical residual than Case 1 and 2. This counter-intuitive phenomenon is probably due to overfitting. In Case 1, the loss function is overwhelmed by noisy measurement terms, potentially causing gradient conflicts during training. The NN over-fits the noise, and sacrifices the generalizability. In Case 2 and 3, with fewer measurement terms, the gradient conflicts might be alleviated.

To conclude, across all test systems and observability levels, M1 consistently outperforms M2, proving that structurally embedding the optimization problem provides more physical consistency than external soft penalties.

\subsection{Performance Comparison Under Severe Contamination Scenarios}\label{sec V-robustness}

\begin{table*}
\centering
\small
\caption{Performance comparison under different noise contamination.}
\label{tab.robustness}
\begin{tabularx}{0.78\textwidth}{c|c|c|ccc|ccc}
\toprule
\specialrule{0em}{0.3pt}{0.5pt}
\toprule
\multirow{2}{*}{\centering Test system} & \multirow{2}{*}{\centering Type} & \multicolumn{4}{c|}{\textbf{M1}} & \multicolumn{3}{c}{\textbf{M2}} \\
\cmidrule(r){3-6} \cmidrule(r){7-9} 
 & & $\bar{\lambda}$ & $\mathcal{L}^{\text{acc}}$ & $\mathcal{L}^{\text{huber}}$ & $\mathcal{L}^{\text{reg}}$ & $\mathcal{L}^{\text{acc}}$ & $\mathcal{L}^{\text{huber}}$ & $\mathcal{L}^{\text{reg}}$ \\
\midrule
\multirow{4}{*}{\centering \textbf{WSCC-9}} & Type 1 & 0.9999 & 0.00086 & 2e-6 & 0.00004 & 0.00157 & 0.09892 & 0.00170 \\
 & Type 2 & 1 & 0.00087 & $<$1e-6 & 0.00003 & 0.00221 & 0.09909 & 0.00145 \\
 & Type 3 & 1 & 0.00117 & 0.00041 & 0.00005 & 0.00195 & 0.08661 & 0.00526 \\
 & Type 4 & 1 & 0.00124 & 0.00039 & 0.00006 & 0.00191 & 0.08620 & 0.00492 \\
\midrule
\multirow{4}{*}{\centering \textbf{IEEE-14}} & Type 1 & 1 & 0.00117 & 1e-6 &	0.00009 & 0.00152 & 0.03235 & 0.00209 \\
 & Type 2 & 1 & 0.00121 & 0.00052 & 0.00034 & 0.00159 & 0.03245 & 0.00245 \\
 & Type 3 & 1 & 0.00120 & 0.00114 & 0.00128 & 0.00212 & 0.03260 & 0.00216 \\
 & Type 4 & 1 & 0.00119 & 0.00051 & 0.00005 & 0.00155 & 0.03235 & 0.00199 \\
\midrule
\multirow{4}{*}{\centering \textbf{Baran 33}} & Type 1 & 1 & 0.00233 & 0.00013 & 0.00705 & 0.00070 & 0.09943 & 0.02501 \\
 & Type 2 & 0.9999 & 0.00462 & 0.00649 & 0.01205 & 0.01470 & 0.13166 & 0.02754 \\
 & Type 3 & 0.9999 & 0.03482 & 0.02353 & 0.06459 & 0.03760 & 0.14717 & 0.01819 \\
 & Type 4 & 0.9999 & 0.07527 & 0.08434 & 0.08585 & 0.04001 & 0.16905 & 0.02153 \\
\midrule
\multirow{4}{*}{\centering \textbf{New England 39}} & Type 1 & 0.9998 & 0.00338 & 0.00003 & 0.00051 & 0.00251 & 0.74316 & 0.05581 \\
 & Type 2 & 0.9971 & 0.00330 & 0.03147 & 0.00274 & 0.00256 & 0.74941 & 0.05490 \\
 & Type 3 & 0.9976 & 0.00342 & 0.01972 & 0.00225 & 0.00304 & 0.75246 & 0.04717 \\
 & Type 4 & 0.9976 & 0.00362 & 0.01877 & 0.00196 & 0.00855 & 0.77452 & 0.04657 \\
\midrule
\multirow{4}{*}{\centering \textbf{IEEE-57}} & Type 1 & 1 & 0.00077 & 3e-6 &	0.00003 & 0.00163 & 0.03697 & 0.00281 \\
 & Type 2 & 0.9998 & 0.00121 & 0.00206 & 0.00139 & 0.00145 & 0.03695 & 0.00337 \\
 & Type 3 & 0.9998 & 0.00121 & 0.00197 & 0.00077 & 0.00176 & 0.03700 & 0.00319 \\
 & Type 4 & 0.9997 & 0.00116 & 0.00237 & 0.00435 & 0.00218 & 0.03786 & 0.00278 \\
\midrule
\multirow{4}{*}{\centering \textbf{Mantovani 136}} & Type 1 & 0.9999 & 0.11526 & 0.00682 & 0.00448 & 0.08390 & 0.08750 & 0.01339 \\
 & Type 2 & 0.9999 & 0.18183 & 0.08024 & 0.00555 & 0.04921 & 0.10967 & 0.02732 \\
 & Type 3 & 0.9999 & 0.21570 & 0.15739 & 0.03898 & 0.35741 & 0.09373 & 0.01228 \\
 & Type 4 & 0.9999 & 0.43567 & 0.15150 & 0.00577 & 0.19199 & 0.08212 & 0.02142 \\
\midrule
\multirow{4}{*}{\centering \textbf{IEEE-300}} & Type 1 & 0.9928 & 0.12615 & 0.00353 & 0.00588 & 0.04873 & 0.42772 & 0.02544 \\
 & Type 2 & 0.9925 & 0.23386 & 0.01859 & 0.01821 & 0.21359 & 0.50707 & 0.02345 \\
 & Type 3 & 0.9922 & 0.32659 & 0.28464 & 0.08257 & 0.04756 & 0.41747 & 0.02537 \\
 & Type 4 & 0.9921 & 0.37718 & 0.30825 & 0.07730 & 0.05145 & 0.44187 & 0.03261 \\
\midrule
\multirow{4}{*}{\centering \textbf{PEGASE 1354}} & Type 1 & 0.9958 & 0.19293 & 0.27864 & 0.09449 & 0.58522 & 0.58723 & 0.14185 \\
 & Type 2 & 0.9954 & 0.34548 & 0.51219 & 0.16782 & 0.56510 & 0.58696 & 0.00952 \\
 & Type 3 & 0.9943 & 0.44367 & 0.62894 & 0.28963 & 0.08642 & 0.53076 & 0.12253 \\
 & Type 4 & 0.9945 & 0.45452 & 0.63260 & 0.29657 & 0.08187 & 0.53252 & 0.12741 \\
\bottomrule
\specialrule{0em}{0.5pt}{0.3pt}
\bottomrule
\end{tabularx}
\end{table*}

\subsubsection{Simulation Settings} 
While previous simulations assumed a fixed outlier penetration rate of $\eta=15\%$, a real-world measurement dataset might exhibit more severe heavy-tailed error distributions, and could encounter structured measurement errors such as remote terminal unit (RTU) communication failures, which is a typical SCADA corruption. To comprehensively simulate these conditions, we designed four distinct types of noise contamination in the measurement dataset:
\begin{itemize}
    \item \textit{Type 1 (Baseline):} Basic Gaussian noise combined with an outlier penetration rate of $\eta=15\%$. As discussed in Section \ref{sec V-setup}, this combination mathematically constitutes a Gaussian Mixture Model, which inherently produces a heavy-tailed distribution, serving as a standard representation of gross errors.
    \item \textit{Type 2 (Severe Heavy-Tailed):} Basic Gaussian noise combined with a doubled outlier penetration rate of $\eta=30\%$.
    \item \textit{Type 3 (Structured Error):} Type 1 combined with RTU communication failures. Specifically, we randomly select 15\% of the buses and force all their associated measurements to exactly zero. This represents a highly structured, non-Gaussian data corruption pattern.
    \item \textit{Type 4 (Compound Error):} Type 2 combined with the 15\% RTU communication failures.
\end{itemize}

We comprehensively tracked the eigenvalue ratio $\bar{\lambda}$ and the three loss metrics ($\mathcal{L}^{\text{acc}}$, $\mathcal{L}^{\text{huber}}$, $\mathcal{L}^{\text{reg}}$) for both our proposed framework (M1) and the PINN baseline (M2) across all the 8 test systems. The detailed statistical results are presented in Table \ref{tab.robustness}.

\subsubsection{Results Analysis}
When the outlier penetration rate $\eta$ doubles from 15\% to 30\% (from Type 1 to 2, or Type 3 to 4), the performance degradation of M1 and M2 is remarkably marginal, demonstrating the robustness of data-driven methods against the heavy-tailed errors. Notably, for small to medium systems (WSCC-9 to IEEE-57), the measurement residuals ($\mathcal{L}^{\text{huber}}$) of M1 remain strictly in the order of $10^{-6}$ to $10^{-3}$. In contrast, the residual of M2 falls in the order of $10^{-2}$ to $10^{-1}$. 

The introduction of RTU communication failures in Type 3 and 4 introduces a fundamentally different challenge. Partial observability discussed in Section \ref{sec V-observe} implies \textit{missing} data (reducing the number of constraints), whereas an RTU error records a false 0 (introducing a severely corrupted constraint). For small and medium topologies, M1 handles these structured errors exceptionally well, maintaining highly accurate and physically consistent estimations. However, as the topology transitions to large-scale grids, the impact of RTU errors becomes more pronounced, leading to an inevitable increase in both $\mathcal{L}^{\text{acc}}$ and $\mathcal{L}^{\text{huber}}$. 

Overall, the estimation performance of M1 still outperforms PINN-based M2 under different settings of noise penetration. This highlights the structural advantage of the $\mathrm{Opt-Layer}$. By treating measurement weights as learnable parameters, M1 could actively identify the outliers.

Interestingly, unlike the partial observability scenarios where $\bar{\lambda}$ noticeably degraded due to the lack of boundary conditions, $\bar{\lambda}$ remains stable across all four noise types in this set of simulation. This phenomenon can be explained from the perspective of measurement redundancy. In these four scenarios, the \textit{quantity} of the measurements remains unchanged (complete observability is structurally maintained); only the \textit{quality} is corrupted. The massive redundancy of the measurement equations, combined with the regularization, continues to bound the feasible region. Therefore, while severe noise pushes the estimated states away from the ground-truth state (increasing $\mathcal{L}^{\text{acc}}$), the relaxed solution space remains tight around the rank-1 physical manifold (maintaining a high $\bar{\lambda}$). This further corroborates our conclusion that in the RSE context, the $\mathrm{Opt-Layer}$ realizes effective denoising and ensures physical consistency.

\begin{table}
\centering
\footnotesize
\caption{Training time (s/epoch) comparison.}
\label{tab.train time}
\begin{tabularx}{\columnwidth}{c|c|c|c|c}
\toprule
\specialrule{0em}{0.3pt}{0.5pt}
\toprule
\textbf{Model} & \textbf{WSCC-9} & \textbf{IEEE-14} & \textbf{Baran 33} & \textbf{New England 39} \\
\midrule
M1 & 3.97 &	7.18 &	20.57 &	26.49	 \\
M2 & 0.76 &	1.31 &	1.78 &	2.37	 \\
M4 & 0.81 &	1.42 &	1.97 &	2.61	\\
\midrule
\textbf{Model} & \textbf{IEEE-57} & \textbf{Mantovani 136} & \textbf{IEEE-300} & \textbf{PEGASE 1354} \\
\midrule
M1 & 47.32 &	136.66 &	427.14 &	2882.9 \\
M2 & 4.12 &	6.76 &	19.01 &	93.16 \\
M4 & 4.16 &	6.88 &	18.36 &	88.39  \\
\bottomrule
\specialrule{0em}{0.5pt}{0.3pt}
\bottomrule
\end{tabularx}
\end{table}

\begin{table}
\centering
\footnotesize
\caption{Inference time (ms/sample) comparison.}
\label{tab.infer time}
\begin{tabularx}{\columnwidth}{c|c|c|c|c}
\toprule
\specialrule{0em}{0.3pt}{0.5pt}
\toprule
\textbf{Model} & \textbf{WSCC-9} & \textbf{IEEE-14} & \textbf{Baran 33} & \textbf{New England 39} \\
\midrule
M1 & 0.81 &	1.17 &	5.5 &	6.76 \\
M2 & 0.04 &	0.02 &	0.02 &	0.02 \\
M4 & 0.06 &	0.06 &	0.07 &	0.07 \\
M8 & 13.11 &	15.5 &	20.85 &	45.59 \\
M9 & 31.4 &	58.23 &	92.81 &	131.51 \\
\midrule
\textbf{Model} & \textbf{IEEE-57} & \textbf{Mantovani 136} & \textbf{IEEE-300} & \textbf{PEGASE 1354} \\
\midrule
M1 & 18.9 &	37.54 &	78.35 &	347.95 \\
M2 & 0.01 &	0.05 &	0.07 &	0.12 \\
M4 & 0.07 &	0.11 &	0.1 &	0.23 \\
M8 & 93.51 &	167.3 &	800.33 &	3624.89 \\
M9 & 230.82 &	514.89 &	1253.92 &	5736.17 \\
\bottomrule
\specialrule{0em}{0.5pt}{0.3pt}
\bottomrule
\end{tabularx}
\end{table}

\begin{table}
\centering
\footnotesize
\caption{Training memory occupation (MB) comparison.}
\label{tab.train mem}
\begin{tabularx}{\columnwidth}{c|c|c|c|c}
\toprule
\specialrule{0em}{0.3pt}{0.5pt}
\toprule
\textbf{Model} & \textbf{WSCC-9} & \textbf{IEEE-14} & \textbf{Baran 33} & \textbf{New England 39} \\
\midrule
M1 & 20.63 & 21.29 & 20.5 & 22.68 \\
M2 & 23.22 & 24.78 & 28.31 & 30.73\\
M4 & 27.79 & 36.26 & 51..87 & 61.71\\
\midrule
\textbf{Model} & \textbf{IEEE-57} & \textbf{Mantovani 136} & \textbf{IEEE-300} & \textbf{PEGASE 1354} \\
\midrule
M1 & 27.2 & 42.41 & 127.49 & 1426.95\\
M2 & 35.64 & 60.45 & 172.08 & 2288.07\\
M4 & 88.88 & 153.75 & 389.83 & 3270.4\\
\bottomrule
\specialrule{0em}{0.5pt}{0.3pt}
\bottomrule
\end{tabularx}
\end{table}

\begin{table}
\centering
\footnotesize
\caption{Inference memory occupation (MB) comparison.}
\label{tab.infer mem}
\begin{tabularx}{\columnwidth}{c|c|c|c|c}
\toprule
\specialrule{0em}{0.3pt}{0.5pt}
\toprule
\textbf{Model} & \textbf{WSCC-9} & \textbf{IEEE-14} & \textbf{Baran 33} & \textbf{New England 39} \\
\midrule
M1 & 20.61 & 20.71 & 20.52 & 21.74 \\
M2 & 22.96 & 24.25 & 27.12 & 29.19 \\
M4 & 24 & 28.78 & 37.42 & 43.5 \\
\midrule
\textbf{Model} & \textbf{IEEE-57} & \textbf{Mantovani 136} & \textbf{IEEE-300} & \textbf{PEGASE 1354} \\
\midrule
M1 & 24.64 & 34.59 & 85.65 & 957.48 \\
M2 & 33.74 & 55.5 & 138.2 & 1433.4 \\
M4 & 59.01 & 96.48 & 231.29 & 1158.77 \\
\bottomrule
\specialrule{0em}{0.5pt}{0.3pt}
\bottomrule
\end{tabularx}
\end{table}

\begin{figure}
    \centering
    \includegraphics[width=\linewidth]{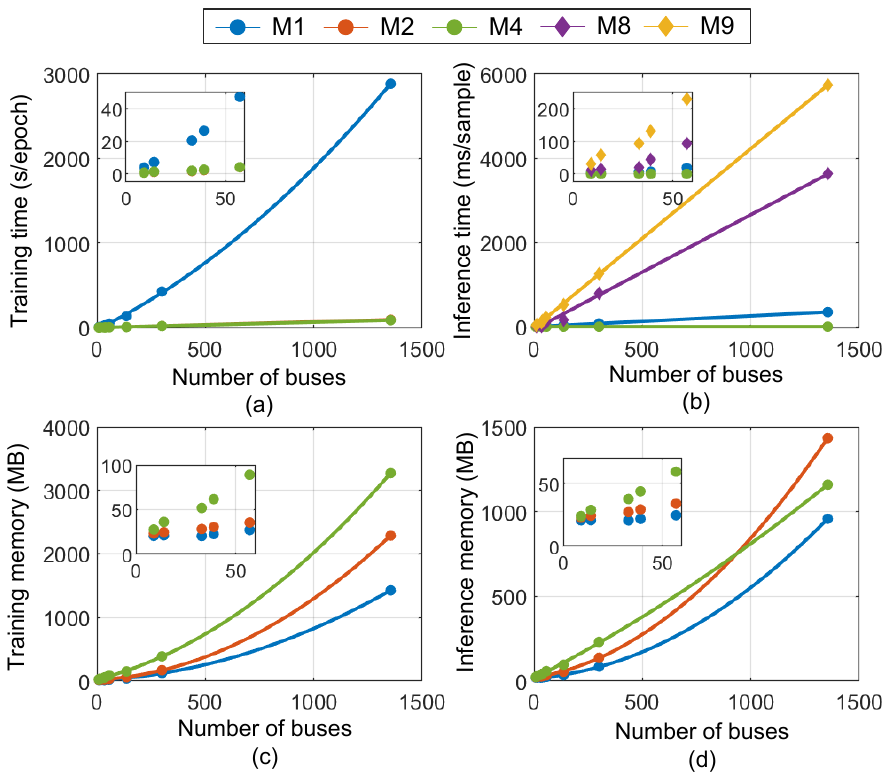}
    \caption{Computational metrics comparison.}
    \label{fig:comp}
\end{figure}

\subsection{Computational Complexity Analysis}\label{sec V-comp}
To analyze and compare the computational complexity of the models introduced in Section \ref{sec V-setup}, both theoretical and empirical complexity analyses are presented in this subsection. 

\subsubsection{Theoretical complexity analysis}
Recall that, $N$ denotes the number of buses in the testing system. Assuming an ideal, fully observable measurement dataset, the number of measurements $M$ is proportional to $N$ (i.e., $\mathcal{O}(N)$).

\textit{a) Complexity of M2-M7:} For PINNs and pure data-driven methods (M2-M7), the computational burden mainly stems from the matrix multiplications during the FP and BP. Since the hidden layer width is proportional to the input dimension $\mathcal{O}(N)$, the complexity of a single NN layer is approximately $\mathcal{O}(N^2)$. Although PINNs introduce additional physical equations into the loss function during training, these algebraic evaluations only require $\mathcal{O}(N)$ operations. Therefore, the overall computational complexity for both training and online inference of M2-M7 is $\mathcal{O}(N^2)$. However, it is worth noting that modern deep learning frameworks execute these dense matrix multiplications in a highly parallelized manner on GPUs. Consequently, the empirical execution time might not strictly exhibit quadratic growth.

\textit{b) Complexity of M1:} The computational complexity of the proposed framework (M1) consists of the FP (solving the SOCP-based RSE problem) and the BP (differentiating the SOCP). During FP, the number of active decision variables in the relaxed SOCP is reduced from $\mathcal{O}(N^2)$ to $\mathcal{O}(N)$ by the sparsification strategy introduced in Section IV-D. The worst-case complexity for FP is $\mathcal{O}(N^3)$. During BP, computing the derivative of the HSDS \eqref{eq diff HSDS} contributes to the majority of the complexity. The auxiliary matrix $\Omega$ has a dimension of $\mathcal{O}(N) \times \mathcal{O}(N)$. Overall, the worst theoretical complexity of the Opt-Layer is $\mathcal{O}(N^3)$ for both training and inference.

\textit{c) Complexity of M8-M9:} For conventional WLAV-based RSE under the AC power flow formulation (M8), the problem is a non-convex nonlinear programming (NLP). Since it is solved via SLP, it requires calculating the Jacobian matrix to linearize the system and solving a LP problem at each iteration. Solving an LP via interior-point methods takes $\mathcal{O}(N^3)$ in the worst case. Furthermore, due to the non-convexity, the number of successive LP iterations required for convergence is unpredictable, making the total computational burden heavier. For conventional WLAV-based RSE under the relaxed SOCP power flow formulation (M9), the worst complexity is similar to the FP of our Opt-Layer $\mathcal{O}(N^3)$.

\subsubsection{Empirical complexity analysis}
We empirically evaluated and compared the training time (s/epoch), inference time (ms/sample), training GPU memory usage (MB) and inference GPU memory usage (MB) across our M1, PINN-based M2 and M4, and optimization-based M8 and M9, as detailed in Tables \ref{tab.train time}-\ref{tab.infer mem} and visualized in Fig. \ref{fig:comp}.

As shown in Table \ref{tab.train time} and Fig. \ref{fig:comp}(a), incorporating the $\mathrm{Opt-Layer}$ (M1) inevitably introduces additional offline training time compared to PINNs (M2, M4). The empirical scalability curves indicate that the training time of M1 grows \textit{quadratically} with the number of buses, whereas M2 and M4 exhibit a near-linear growth. This stems from the need to solve RSE problems during FP and perform implicit differentiation during BP when applying M1. It is important to highlight that although the theoretical worst-case complexity of M1 is $\mathcal{O}(N^3)$, the empirical training time exhibits an approximately $\mathcal{O}(N^2)$ growth. This significant reduction is attributed to the inherent sparsity of power grid topologies. For M2 and M4, although the theoretical complexity is around $\mathcal{O}(N^2)$, the empirical complexity enjoys a near-linear growth. This is because of the parallel computing acceleration on GPUs. While differentiable optimization layers can leverage batch processing on GPUs (as introduced in \cite{amos2017optnet}), these underlying algebraic operations possess inherent sequential data dependencies. Consequently, M1 cannot saturate GPU cores as efficiently as standard NNs, which remains a common bottleneck in the field of differentiable optimization. \textit{However, under the "offline training, online deployment" paradigm, this computational burden is strictly confined to the offline phase}, where time constraints are generally flexible.

The online inference time is the decisive metric for real-time applications. As demonstrated in Table \ref{tab.infer time} and Fig. \ref{fig:comp}(b), while M1's inference time is longer than that of M2 and M4, it still scales favorably with an acceptable slope. More importantly, M1 is \textit{remarkably faster} than the traditional optimization-based models (M8, M9), and could achieve near-instantaneous decision-making even for large-scale grids. This facilitates the real-time application of M1 in the context of RSE.

Tables \ref{tab.train mem}, \ref{tab.infer mem} and Fig. \ref{fig:comp}(c), (d) track the memory usage. Interestingly, while the memory consumption for M1, M2, and M4 all exhibit a quadratic relationship with the number of buses, M1 consistently consumes less memory than M2 and M4 during both training and inference. This advantage stems from the structural efficiency of the proposed framework. The $\mathrm{Opt-Layer}$ inherently captures the complex physical manifold, allowing the NN to maintain a much \textit{shallower architecture} compared to PINNs \cite{amos2017optnet}.

To conclude, an acceptable increase in offline training time is traded in M1 for a physically consistent and real-time capable robust state estimator. As evidenced by the measurement residuals being orders of magnitude lower than those of PINNs, this trade-off yields significantly superior physical consistency that PINN methods simply cannot achieve.

\section{Conclusion}\label{sec6}
\subsection{Major conclusions}
RSE is a critical tool for monitoring the operation states of smart grids considering the presence of outliers in measurements. Traditional E2E learning approaches for RSE prioritize regression accuracy while being hard to ensure the physical consistency. To bridge this gap, this paper develops a novel optimization-embedded E2E learning framework. The core innovation lies in the construction of a differentiable optimization layer $\mathrm{Opt-Layer}$ strictly embedded with KKT optimality conditions. To calculate the derivatives of global optimal solution with respect to problem parameters according to KKT conditions, convex relaxation is utilized. Extensive simulations under heterogeneous scenarios and settings have validated the effectiveness of the proposed framework in boosting the overall estimation quality, including significantly lower measurement residuals compared with classical E2E learning-based approaches.

The proposed framework offers distinct advantages and also presents avenues for future research. A primary strength is the rigorous optimality conditions embedding of $\mathrm{Opt-Layer}$ during BP. This leads to the lowest measurement residuals compared to existing mainstream learning methods, as demonstrated in Section \ref{sec5}. Through treating measurement weights as learnable parameters, the proposed framework exhibits distinguished robustness even under severe noise contamination. Furthermore, this architecture allows the NN to achieve enhanced richness of representation, maintaining a shallower structure. However, applying this centralized framework to ultra-large-scale systems presents specific limitations. First, since the FP involves repeatedly solving optimization problems, the offline training process can be computationally demanding. Second, the exactness of the convex relaxation in the embedded optimization layer may degrade as the system size grows or under extremely low observability. Fortunately, the learned prior knowledge of the NN effectively compensates for this inexactness, ensuring physical consistency, while the rapid inference speed facilitates the real-time deployment of our framework.

\subsection{Future Work}
Future work could include advanced acceleration techniques to reduce the computational burden of training. The research on dataset representativeness is also worth in-depth investigation. Moreover, extending the framework to a distributed or multi-agent architecture would enhance its scalability and training efficiency for ultra-large-scale power systems. Furthermore, extending our framework to topology-aware dynamic RSE through graph-based learning, and considering the uncertainties of power network parameters (e.g., line resistance and reactance) through joint state and parameter estimation also serve as promising research topics for enhanced estimation robustness. In addition, establishing formal theoretical robustness bounds, convex relaxation exactness under the learning dynamics and global convergence properties of NNs embedded with optimization layers remain open challenges and insightful directions for future theoretical research.

\appendices
\section{Recommended Observability and Operation Conditions}
While the SOCP relaxation is theoretically exact for radial networks \textit{under ideal noiseless conditions}, achieving an approximated rank-1 solution for general power systems in the \textit{noisy} RSE context relies on the following recommended conditions \cite{zhang2017conic}:
\begin{enumerate}
    \item Observability condition: Voltage magnitude measurements $V_i$ are available for all nodes. Active power flow measurements $P_{ij}$ are available for a set of branches that forms at least one spanning tree of the power network.
    \item Operation condition: The phase angle difference across any line is maintained within the range of -90° to 90°.
\end{enumerate}

These conditions could be generally satisfied in modern power systems equipped with SCADA, operating under normal (i.e., non-heavily loaded) conditions \cite{korres2011state}. While they do not strictly guarantee exactness for large-scale meshed grids, they serve as the fundamental prerequisites to tighten the feasible region.

\section{Proof of Corollary 1}
Recall that, the total active power loss of a power system could be expressed as the algebraic addition of all nodal active power injection, namely $P_{loss}=\sum_{i\in \mathcal{N}}P_i$. According to AC power flow equations, the complex power injection at node $i$ is calculated as:
\begin{equation}
    \textstyle S_i=V_iI_i^*=V_i(\sum_{j\in \mathcal{N}}Y_{ij}V_j)^*
\end{equation}
whose real part denotes the active power injection at node $i$:
\begin{equation}
    \textstyle P_i=\mathrm{Re}(S_i)=\mathrm{Re}(\sum_{j\in \mathcal{N}}Y_{ij}^*V_iV_j^*)
\end{equation}

Given the off-diagonal elements of the introduced auxiliary matrix $X_{ij}=V_iV_j(\cos \theta_{ij}+j\sin \theta_{ij})=V_iV_j^*$, the total active power loss is therefore modeled as follows:
\begin{equation}
     \textstyle P_{loss}=\mathrm{Re}(\sum_{i\in \mathcal{N}}\sum_{j\in \mathcal{N}}Y_{ij}^*X_{ij})
\end{equation}

We may also let $\Phi=(Y^*)^{\top}\bm{X}$, whose diagonal elements can be expressed as:
\begin{equation}
    \textstyle \Phi_{jj} = \sum_{i\in \mathcal{N}}(Y^*_{ji})^{\top}X_{ij} = \sum_{i\in \mathcal{N}}Y^*_{ij}X_{ij}
\end{equation}

Summing all off-diagonal elements, we have the trace of $\Phi$:
\begin{equation}
    \textstyle \mathrm{Tr}(\Phi)= \sum_{j\in \mathcal{N}}\Phi_{jj}=\sum_{i\in \mathcal{N}}\sum_{j\in \mathcal{N}}Y_{ij}^*X_{ij}
\end{equation}
whose real part is exactly the total active power loss of the power system, namely:
\begin{equation}
    P_{loss}=\mathrm{Re}(\mathrm{Tr}(\Phi))=G_{ij}\mathrm{Re}(X_{ij}) + B_{ij}\mathrm{Im}(X_{ij})
\end{equation}
which is exactly the regularization term in the objective of $\mathrm{(P2)}$ and $\mathrm{(P3)}$. It is also a linear function of decision variables.

\ifCLASSOPTIONcaptionsoff
  \newpage
\fi

\bibliographystyle{IEEEtran}
\bibliography{IEEEabrv,Bibliography}

\vfill

\end{document}